\documentclass[12pt,letterpaper]{article}
\usepackage{epsfig,rotating,setspace,latexsym,amsmath,epsf,amssymb,bm,theorem}
\usepackage{cite,appendix,psfrag,subfigure,algorithmic}
\usepackage{amsfonts}

\title{Wiretap Channels: Implications of the More Capable Condition and Cyclic Shift Symmetry\thanks{This work was supported by NSF Grants CCF 07-29127, CNS 09-64632, CCF 09-64645, CCF 10-18185 and presented in part at the IEEE International Symposium on Information Theory (ISIT), St. Petersburg, Russia, July 2011 and at the Allerton Conference on Communications, Control and Computing, Monticello, IL, September 2011.}}

\author{Omur Ozel \qquad Sennur Ulukus \\
\normalsize Department of Electrical and Computer Engineering\\
\normalsize University of Maryland, College Park, MD 20742 \\
\normalsize {\it omur@umd.edu} \qquad {\it ulukus@umd.edu}}

\newcommand{\ev}{{\mathbf{e}}}
\newcommand{\pv}{{\mathbf{p}}}
\newcommand{\qv}{{\mathbf{q}}}

\newcommand{\uv}{{\mathbf{u}}}

\newtheorem{theorem}{Theorem}

\newtheorem{corollary}{Corollary}
\newtheorem{definition}{Definition}
\newtheorem{lemma}{Lemma}

\newenvironment{Proof}[1]{\medskip\par\noindent{\bf Proof:\,}\,#1}{{\mbox{\,$\blacksquare$}\par}}

\begin{document}

\maketitle

\begin{abstract}
Characterization of the rate-equivocation region of a general wiretap channel involves
two auxiliary random variables: $U$, for rate splitting and $V$, for channel prefixing. Evaluation of regions involving auxiliary random variables is generally difficult. In this paper, we explore specific classes of wiretap channels for which the expression and evaluation of the rate-equivocation region are simpler. In particular, we show that when the main channel is more capable than the eavesdropping channel, $V=X$ is optimal and the boundary of the rate-equivocation region can be achieved by varying $U$ alone. Conversely, we show under a mild condition that if the main receiver is not more capable, then $V=X$ is strictly suboptimal. Next, we focus on the class of cyclic shift symmetric wiretap channels. We explicitly determine the optimal selections of rate splitting $U$ and channel prefixing $V$ that achieve the boundary of the rate-equivocation region. We show that optimal $U$ and $V$ are determined via cyclic shifts of the solution of an auxiliary optimization problem that involves only one auxiliary random variable. In addition, we provide a sufficient condition for cyclic shift symmetric wiretap channels to have $U=\phi$ as an optimal selection. Finally, we apply our results to the binary-input cyclic shift symmetric wiretap channels. We solve the corresponding constrained optimization problem by inspecting each point of the $I(X;Y)-I(X;Z)$ function. We thoroughly characterize the rate-equivocation regions of the BSC-BEC and BEC-BSC wiretap channels. In particular, we find that $U=\phi$ is optimal and the boundary of the rate-equivocation region is achieved by varying $V$ alone for the BSC-BEC wiretap channel.
\end{abstract}

\newpage

\section{Introduction}

\begin{figure}[t]
\centering
\includegraphics[width=0.47\linewidth]{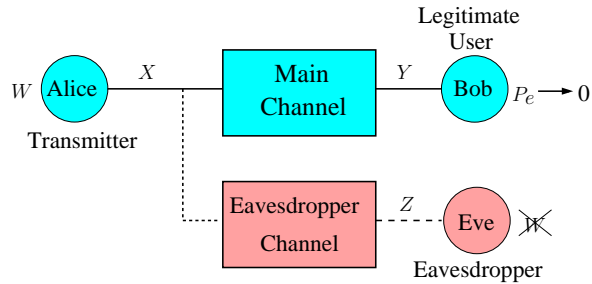}
\caption{The wiretap channel.}
\label{sys}
\end{figure}

We consider the discrete memoryless wiretap channel shown in
Fig.~\ref{sys}. The capacity region of this channel is
characterized by the rate, $R$, between the legitimate users
Alice and Bob, and the equivocation, $R_e$, at the eavesdropper
Eve. Wyner \cite{wyner78} characterized the rate-equivocation
region when the received signal at Eve is a degraded version of
the signal received at Bob. Csisz\'{a}r and K\"{o}rner
\cite{CK78} characterized the rate-equivocation region for
general, not necessarily degraded, wiretap channels.

Csisz\'{a}r and K\"{o}rner's characterization involves two
auxiliary random variables: $U$, for rate splitting, and $V$,
for channel prefixing. Evaluation of capacity regions involving
auxiliary random variables is generally difficult, and it is
desirable to determine cases where the auxiliary random variables are either not needed or their optimal selection is simplified. For the wiretap channel, under certain conditions, it is known that the use of one or both of these
auxiliary random variables is unnecessary. For instance, if the
wiretap channel is degraded, neither rate splitting nor channel
prefixing is necessary, i.e., the selection $U=\phi$ and $V=X$
is optimal, for the entire rate-equivocation region
\cite{wyner78}. In fact, the same conclusion holds if the
wiretap channel is less noisy \cite[Theorem 3]{CK78}. For general wiretap
channels, for the purposes of characterizing the {\it secrecy
capacity}, i.e., the largest equivocation, rate splitting is
unnecessary, i.e., $U=\phi$ is optimal \cite{CK78}; further, if
the wiretap channel is more capable, then channel prefixing as
well is unnecessary, i.e., $U=\phi$ and $V=X$ are optimal \cite{CK78}.

\begin{figure}[t]
\centering
\includegraphics[width=0.44\linewidth]{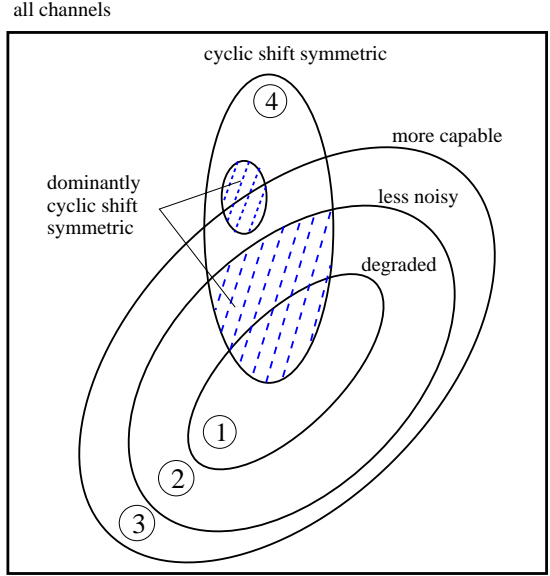}
\caption{Inclusion relations among the classes of wiretap channels.}
\label{sch}
\end{figure}

In this paper, we explore specific classes of wiretap channels for which calculation of the optimal rate
splitting and/or channel prefixing parameters is simpler. The inclusion relations among the classes of wiretap channels
considered in this paper are shown in Fig.~\ref{sch}. First, we
show that if the wiretap channel is more capable, then channel
prefixing is unnecessary; that is, the rate-equivocation region can be
characterized by rate splitting, i.e., $V=X$ is optimal and the
boundary of the rate-equivocation region can be traced with
optimal $(U,X)$ only. Conversely, we prove under a mild condition that, if the channel
is not more capable, then channel prefixing is strictly
necessary, i.e., $V \neq X$ is strictly needed.

Next, we study the class of cyclic shift symmetric wiretap channels. We explicitly determine the optimal selection of $U$ and $V$ that achieve the boundary of the rate-equivocation region. In particular, optimal $U$ and $V$ are expressed in terms of the cyclic shifts of the solution of an auxiliary optimization problem that involves only one auxiliary random variable. This is a considerable reduction in the computation requirement for the calculation of (the boundary of) the rate-equivocation region. We provide the cardinality bound on this single auxiliary random variable appearing in the optimization problem. Then, we formulate the problem as a constrained optimization problem. We provide a sufficient condition under which rate splitting is unnecessary, i.e., $U=\phi$ is optimal and the boundary of the rate-equivocation region is obtained by varying $V$ alone. In particular, we show that if $I(X;Y)-I(X;Z)$ is maximized at the uniform distribution, i.e., if the channel is dominantly cyclic shift symmetric, then this sufficient condition is satisfied and hence rate splitting is unnecessary. Moreover, we show that if the main channel is more capable and both channels are cyclic shift symmetric, then $(C_B,C_s)$ rate-equivocation pair is achievable. We also discuss an extension of the notion of cyclic shift symmetry for continuous alphabets. Finally, we apply our results to the binary-input cyclic shift symmetric wiretap channels. We investigate two examples that illustrate the considered cases: BSC-BEC and BEC-BSC wiretap channels. We provide full characterizations for the rate-equivocation regions of the BSC-BEC and BEC-BSC wiretap channels. We identify the conditions on the BSC cross-over probability and the BEC erasure probability that guarantee that channel prefixing or rate splitting or both are unnecessary. In particular, we find that rate splitting is never necessary for the BSC-BEC wiretap channel. 

\section{Model and Background}
As in Fig.~\ref{sys}, Alice communicates with Bob in the
presence of an eavesdropper, Eve. The input and output
alphabets, $\mathcal{X}$, $\mathcal{Y}$ and $\mathcal{Z}$, are
finite. The main channel is characterized by $p(y|x)$ and has
capacity $C_B=\max_{P_x} I(X;Y)$. Similarly the wiretapper
channel is characterized by $p(z|x)$ and has capacity
$C_E=\max_{P_x} I(X;Z)$. $W$ represents the message to be sent
to Bob and kept secret from Eve with $W \in
\mathcal{W}=\{1,\ldots,2^{nR}\}$. Alice uses an encoder
$\varphi: \mathcal{W} \rightarrow \mathcal{X}^n$ to map each
message to a channel input of length $n$. Bob uses a decoder
$\psi: \mathcal{Y}^n \rightarrow \mathcal{W}$. The probability of
error is: $P_e=\mbox{Pr}\left[\psi(Y^n)\neq W\right]$. The rate
$R$ is achievable with equivocation $R_e$, if $P_e \rightarrow
0$ as $n \rightarrow \infty$, and
\begin{align}
R_e = \lim_{n\rightarrow \infty} \frac{1}{n} H(W | Z^n)
\end{align}
Perfect secrecy\footnote{We use the weak secrecy notion.
However, for discrete wiretap channels weak and strong secrecy are
equivalent \cite{cszr,maurer}.} is achieved if $\frac{1}{n}I(W;Z^n)
\rightarrow 0$ and the {\it secrecy capacity} $C_s$ is the highest
achievable perfectly secure rate $R$. The maximum possible
equivocation is also $C_s$.

The input distribution $P_x$
belongs to the $|\mathcal{X}|$ dimensional probability simplex denoted as
\begin{align}
\Delta = \left\{(p_1,\ldots,p_{|\mathcal{X}|}) \left|
\sum_{i=1}^{|\mathcal{X}|} p_i = 1,\quad p_i \geq 0, ~\forall i \right. \right\}
\end{align}
Throughout the paper, $f_{\mu}(.)$ denotes the following function of
the input distribution $P_x$
\begin{align}
f_{\mu}(P_x) = (\mu + 1)I(X;Y) - I(X;Z)
\end{align}
where $\mu \geq 0$ is an arbitrary parameter. We denote $f_{0}(.)$ simply as $f(.)$. Note that $f_{\mu}(.)$ is continuous and differentiable for all $\mu \geq 0$. 

Csisz\'{a}r and K\"{o}rner \cite{CK78} characterized the entire
rate-equivocation region as stated in the following theorem.
\begin{theorem}[\!\!{\cite[Corollary 2]{CK78}}]
\label{ck_thm}
$(R,R_e)$ pair is in the rate-equivocation region if and only if there
exist $U \rightarrow V \rightarrow X \rightarrow Y,Z$ such that
$I(U;Y)\leq I(U;Z)$, and
\begin{align}
0 \leq R_e & \leq I(V;Y|U) - I(V;Z|U) \label{ck1}\\
R_e \leq R & \leq I(V;Y) \label{ck2}
\end{align}
Further, the secrecy capacity is
\begin{align}
C_s = \max_{V\rightarrow X \rightarrow Y,Z}
I(V;Y) - I(V;Z) \label{ck3}
\end{align} Finally, the cardinality bounds on the alphabets of the auxiliary random variables are
\begin{align} \label{bd1} |\mathcal{U}| &\leq |\mathcal{X}|+3 \\ \label{bd2} |\mathcal{V}| &\leq |\mathcal{X}|^2 + 4 |\mathcal{X}| + 3 \end{align}
\end{theorem}

\begin{figure}[t]
\centering
\includegraphics[width=0.6\linewidth]{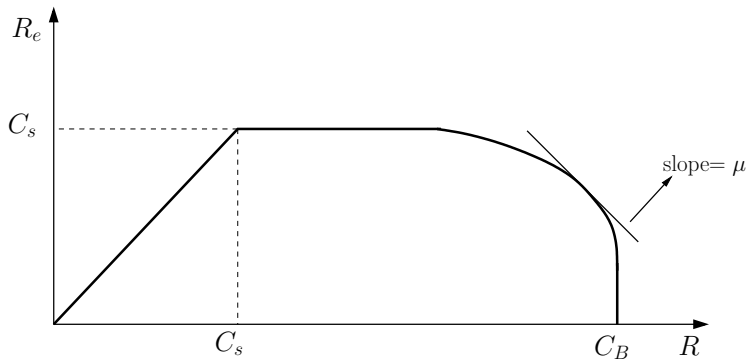}
\caption{ Characterization of the upper right boundary of the rate-equivocation region.}
\label{reg}
\end{figure}

The rate-equivocation region of a wiretap channel is a convex region.
Therefore, the upper right boundary is traced by solving the following
optimization problem for all $\mu\geq 0$ as in Fig. \ref{reg}:
\begin{align}
\label{ree}
\max_{U,V,X} \quad \mu I(V;Y) + I(V;Y|U) - I(V;Z|U)
\end{align}
Note that this optimization problem is
computable due to the bounds on the sizes of
$U$ and $V$ in (\ref{bd1}) and (\ref{bd2}) in Theorem \ref{ck_thm}. In the sequel, we refer to the solution of the optimization problem in (\ref{ree}) as the optimal selections $U^*$, $V^*$ and $X^*$. These optimal selections depend implicitly on the value of $\mu$. 
The optimal value of the objective function in (\ref{ree}) at
$\mu=0$ is the secrecy capacity $C_s$. In this case, $U$ is
unnecessary, and in fact, we get (\ref{ck3}) \cite{CK78}. 
Note that the bounds on the cardinalities of $U$ and $V$ in (\ref{bd1})-(\ref{bd2}) in Theorem \ref{ck_thm} are valid in general. However, the specific cardinality bound on $V$ for the optimization problem in (\ref{ree}) when $\mu=0$, or equivalently the problem in (\ref{ck3}), is \begin{align} \label{carx} |\mathcal{V}| \leq |\mathcal{X}| \end{align} In order to prove the cardinality bound in (\ref{carx}), given $V \rightarrow X \rightarrow Y,Z$ with PMFs $p(v)$ and $p(x|v)$, we fix the following $|\mathcal{X}|$ continuous functions of $p(x|v)$: 
\begin{equation}
\label{exppp}
g_j(p_{X|V}(x|v))=\left\{\begin{array}{ll}
p_{X|V}(j|v), & \mbox{$j=1,\ldots,|\mathcal{X}|-1$}\\
I(X;Y|V=v)-I(X;Z|V=v), & \mbox{$j=|\mathcal{X}|$}
\end{array}
\right.
\end{equation}
From Lemma 3 in \cite{ahlswede75} and the strengthened Caretheodory theorem of Fenchel-Eggleston in \cite{salehi78}, we can find another random variable $V^{'}$ with cardinality at most $|\mathcal{X}|$ such that $V^{'} \rightarrow X \rightarrow Y,Z$ and $I(X;Y|V) - I(X;Z|V)=I(X;Y|V^{'})-I(X;Z|V^{'})$ as well as $p(x)=\int p_{X|V}(x|v) dF(v) = \sum_{v^{'}} p_{X|V^{'}}(x|v^{'})p(v^{'})$ for $x=1,\ldots,|\mathcal{X}|-1$; see also Appendix C in \cite{elgamal10}. Since \begin{align} I(V;Y) - I(V;Z) = I(X;Y) - I(X;Z) - [I(X;Y|V) - I(X;Z|V)] \end{align} we conclude that $I(V;Y) - I(V;Z) = I(V^{'};Y) - I(V^{'};Z)$. Therefore, $|\mathcal{V}| \leq |\mathcal{X}|$ cardinality is sufficient to solve the optimization problem in (\ref{ck3}).

\section{More Capable Wiretap Channels}

More capable condition is a partial ordering for discrete
memoryless channels as formally defined below.
\begin{definition}[\!\!\cite{CK78}]
$p(y|x)$ is more capable than $p(z|x)$ if $f(P_x) \geq 0$ for
all $P_x \in \Delta$.
\end{definition}
A wiretap channel is more capable if the main channel is more
capable than the eavesdropping channel.

In {\cite[Theorem 3]{CK78}}, Csisz\'{a}r and K\"{o}rner observe
that if the wiretap channel is more capable, then channel
prefixing is unnecessary, i.e., $V=X$ is optimal, for achieving
the secrecy capacity. We will strengthen this result. We will
prove that if the wiretap channel is more capable, then channel
prefixing is unnecessary for achieving the entire boundary of the rate-equivocation region. Conversely, we will prove, under a mild condition, that if the wiretap channel is not more capable, then $V=X$ is strictly suboptimal, i.e., there exists $V \neq X$ that improves the rate-equivocation region compared to $V=X$.

Let $\ev_j$ denote the elementary PMF where all the mass is
concentrated in the $j$th coordinate, i.e., its $j$th entry is
$1$ and all other entries are zero. Note that $\ev_j$,
$j=1,\ldots,|\mathcal{X}|$, form the canonical basis for the $|\mathcal{X}|$
dimensional Euclidean space, and in particular, $\Delta$ is the
convex hull of $\ev_j$, $j=1,\ldots,|\mathcal{X}|$. An important topological property of $\Delta$ is stated in the next lemma, namely a point in the simplex $\Delta$ partitions the simplex into $|\mathcal{X}|$ sub-simplexes in a specific way. 
\begin{lemma}
\label{lm1} Let $\pv$ and $\pv'$ be two PMFs in $\Delta$.
There exists a PMF $\qv$ and an index set $J \subset
\{1,\ldots,|\mathcal{X}|\}$ with $|J|=|\mathcal{X}|-1$ such that
\begin{align}\label{rep}
\pv' = q_1 \pv + \sum_{i=1}^{|\mathcal{X}|-1}q_{i+1} \ev_{j_i}
\end{align}
where $j_i \in J$ for $i=1,\ldots, |\mathcal{X}|-1$. In particular, $q_1>0$ if $\pv'$ is an interior point of $\Delta$. 
\end{lemma}
\begin{Proof}
The probability simplex $\Delta$ has corner points $\ev_j$,
$j=1,\ldots,|\mathcal{X}|$. Given $\pv \in \Delta$, we can find a
triangulation \cite{triangle} $\{\mathcal{D}_i\}_{i=1}^{|\mathcal{X}|}$ of $\{\ev_1, \ldots, \ev_{|\mathcal{X}|},
\pv\}$ by combining $|\mathcal{X}|-1$ of the corner points and $\pv$. Then, we get
$\Delta=\bigcup_{i=1}^{|\mathcal{X}|} \mathcal{D}_i$ where $\mathcal{D}_i$
is the convex hull of $\left[ \{\ev_1,\ldots,\ev_{|\mathcal{X}|}\} \setminus
\{\ev_i\} \right] \cup \{\pv\}$, $i=1,\ldots,|\mathcal{X}|$. If $\pv$ has a zero entry, then some $\mathcal{D}_i$ has smaller dimensionality, however this does not violate the generality. Hence, a
given PMF $\pv'$ resides inside one of $\mathcal{D}_i$. Moreover, if $\pv'$ has all non-zero entries, then it is not in the convex hull of any proper subset of $\ev_j$, $j=1,\ldots,|\mathcal{X}|$. Hence, $q_1>0$ in this case.  
\end{Proof}

\begin{figure}[t]
\centering
\includegraphics[width=0.4\linewidth]{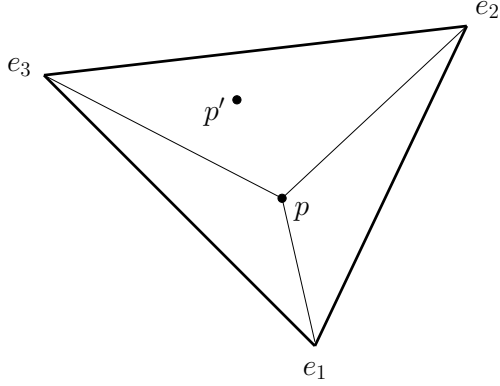}
\caption{Partitioning the probability simplex $\Delta$.}
\label{part}
\end{figure}

Lemma \ref{lm1} says that a point in $\Delta$ partitions it into $|\mathcal{X}|$ sub-simplexes which are convex hulls of $|\mathcal{X}|-1$ of the vertices $\ev_j$ and the point itself. We illustrate this partitioning for $|\mathcal{X}|=3$ in Fig. \ref{part}. As a consequence, any PMF can be expressed as a convex combination of any other PMF and $|\mathcal{X}|-1$ of the $|\mathcal{X}|$ canonical PMFs $\ev_j$. In fact, this partition and hence the representation in (\ref{rep}) is unique. Only the existence of such a representation is sufficient for our arguments in this paper. In particular, we use this existence result to prove the main theorem of this section which is stated next. The proof of this theorem is provided in Appendix \ref{app2}.
\begin{theorem} \label{tm_mc}
If the wiretap channel is more capable, $V^*=X$ is optimal for the entire boundary of the rate-equivocation region and the cardinality bound on $U^*$ is \begin{align} |\mathcal{U^*}| \leq |\mathcal{X}| \end{align} Conversely, if the wiretap channel is not more capable, $V=X$ is strictly suboptimal provided that $f(P_x)$ is maximized at an interior point of $\Delta$.
\end{theorem}

As a result, if the wiretap channel is more capable, channel prefixing is not necessary and hence the computation of the rate-equivocation region is considerably simplified. Moreover, the bound on the necessary rate splitting reduces by 3 compared to Csisz\'{a}r and K\"{o}rner's bound (from $|\mathcal{X}|+3$ to $|\mathcal{X}|$). Another remark is that the direct part in Theorem \ref{tm_mc} immediately extends for continuous alphabet wiretap channels. However, the converse part does not immediately extend as the proof is not valid for infinite dimensional spaces of probability density functions.

We next review less noisy channels for future reference. Less noisy condition is a stronger
partial ordering than more capable condition.
\begin{definition}[\!\!\cite{CK78}]
$p(y|x)$ is less noisy than $p(z|x)$ if $I(U;Y)\geq I(U;Z)$ for all $U \rightarrow
X \rightarrow Y,Z$.
\end{definition}
A wiretap channel is less noisy if the main channel is less
noisy than the eavesdropping channel. If a wiretap channel is less noisy (regions \textcircled{1} and
\textcircled{2} in Fig.~\ref{sch}), neither rate splitting nor
channel prefixing is necessary for the entire rate-equivocation
region \cite[Theorem 3]{CK78}. 

\section{Cyclic Shift Symmetric Wiretap Channels}

In this section, we focus on cyclic shift symmetric channels.
\begin{definition}[\!\!\cite{invariance_ita}]
$p(y|x)$ is cyclic shift symmetric if $I(X;Y)$ is invariant
under cyclic shifts of the input distribution.
\end{definition}
Cyclic shift symmetric channels are an important class that
includes binary symmetric, binary erasure, and type-writer
channels. A key property of cyclic shift symmetric channels is
that the input distribution that maximizes the mutual
information is the uniform distribution {\cite[Theorem
2]{invariance_ita}}. A wiretap channel is cyclic shift
symmetric if both the main channel and the eavesdropping
channel are cyclic shift symmetric.

In the following theorem, we determine the structure of the optimal auxiliary random variables $U^*$ and $V^*$ as well as the channel input $X^*$ for cyclic shift symmetric wiretap channels. Remarkably, the optimizing rate splitting $U^*$ and channel prefixing $V^*$ parameters can be determined by solving an auxiliary optimization problem over only one auxiliary random variable. In addition, the cardinality bounds on $U^*$ and $V^*$ are reduced to $|\mathcal{X}|$ and $|\mathcal{X}|^2$, respectively, compared to the general case in (\ref{bd1}) and (\ref{bd2}). We provide the proof of this theorem in Appendix \ref{app3}.
\begin{theorem} \label{cyc}
In a cyclic shift symmetric wiretap channel, the optimal selection of the auxiliary random variables $U^*$ and $V^*$ in (\ref{ree}) have the cardinalities $|\mathcal{U}^*| \leq |\mathcal{X}|$ and $|\mathcal{V}^*| \leq |\mathcal{X}|^2$, respectively, with the following structure: \begin{align} \label{dd1} p(U^*=u)&=\frac{1}{|\mathcal{X}|}, \qquad u \in \{1,\ldots,|\mathcal{X}|\} \\ p(V^*=(u-1)|\mathcal{X}|+v|U^*=u)&= p(\hat{V}=v), \qquad u,v \in \{1,\ldots,|\mathcal{X}|\} \\ p(V^*=v|U^*=u)&= 0, \quad u \in \{1,\ldots,|\mathcal{X}|\}, v \notin \{(u-1)|\mathcal{X}|+1,\ldots,u|\mathcal{X}|\} \\ p(x|V^*=(u-1)|\mathcal{X}|+v) &= p(x|\hat{V}=v)(u-1), \quad  u,v,x \in \{1,\ldots,|\mathcal{X}|\} \label{dd2}
\end{align}
where $p(X=x|\hat{V}=v)(u-1)$ denotes the $u-1$st cyclic shift of the distribution $p(x|\hat{V}=v)$. Moreover, the distributions $p(\hat{V}=v)$ and $p(X=x|\hat{V}=v)$ with $|\mathcal{\hat{V}}|\leq |\mathcal{X}|$ are the optimizers of the following auxiliary optimization problem: \begin{align} \label{vv} \max_{\hat{V} \rightarrow X \rightarrow Y,Z} \bigg( I(X;Y) - I(X;Z) - \left[(\mu+1)I(X;Y|\hat{V}) - I(X;Z|\hat{V})\right] \bigg)^+ \end{align} where $(x)^+=\max\{0,x\}$.
\end{theorem}

\begin{figure}[t]
\centering
\includegraphics[width=0.7\linewidth]{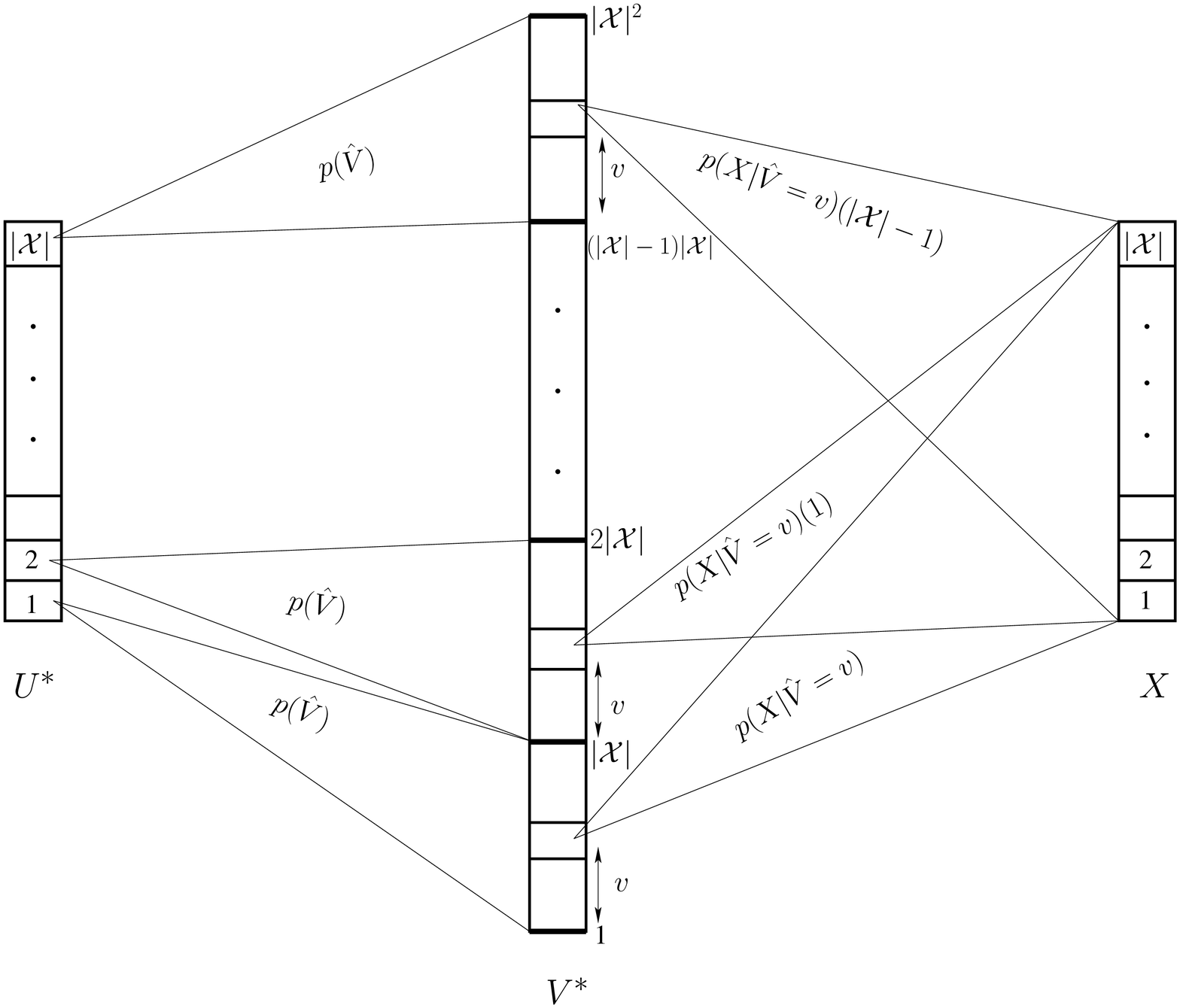}
\caption{The structure of the optimal $U^* \rightarrow V^* \rightarrow X$ for cyclic shift symmetric wiretap channels. $p(\hat{V})$ and $p(X|\hat{V}=v)$, $v \in \{1,\ldots,|\mathcal{X}|\}$ are the solutions of the auxiliary optimization problem in (\ref{vv}).}
\label{uvx}
\end{figure}

We illustrate the specific structure of the optimal auxiliary random variables and the channel input in Fig. \ref{uvx}. In particular, each element of $U^*$ generates the optimizing PMF $p(\hat{V})$ over $|\mathcal{X}|$ elements of $V^*$. The first $|\mathcal{X}|$ elements of $V^*$ generate the optimizing conditional PMF $p(X|\hat{V}=v)$ over $X$. The remaining elements of $V^*$ generate cyclic shifts of $p(X|\hat{V}=v)$ over $X$.    
An equivalent representation for the optimal selections can be obtained by letting $V^*=(V_1^*,V_2^*)$ with $|\mathcal{V}_1^*|=|\mathcal{V}_2^*|=|\mathcal{X}|$:
\begin{align} \label{eqv1} p(U^*=u)&=\frac{1}{|\mathcal{X}|}, \qquad u \in \{1,\ldots,|\mathcal{X}|\} \\ p(V^*=(v_1,v_2)|U^*=u)&= p(\hat{V}=v_1)\delta(v_2-u), \qquad u,v_1,v_2 \in \{1,\ldots,|\mathcal{X}|\} \\ p(x|V^*=(v_1,v_2)) &= p(x|\hat{V}=v_1)(v_2-1), \qquad  v_1,v_2 \in \{1,\ldots,|\mathcal{X}|\} \label{eqv2}
\end{align} Note that $U^*$ is a deterministic function of $V^*$ as stated in \cite[Theorem 1]{CK78}. This is verified easily from the equivalent representation in (\ref{eqv1})-(\ref{eqv2}). Given $V^*=(V_1^*=v_1,V_2^*=v_2)$, $U^*=v_2$ with probability 1. However, $V^*$ is a stochastic function of $U^*$. These can also be verified from Fig. \ref{uvx}. 

The optimization problem in (\ref{vv}) is a constrained optimization problem over $|\mathcal{X}|^2 -1$ variables: $|\mathcal{X}|$ probability distributions on $X$, $p(X=x|\hat{V}=v_i)$. Each probability distribution accounts for $|\mathcal{X}|-1$ variables for $i=1,\ldots,|\mathcal{X}|$. In addition, the distribution for $\hat{V}$ accounts for $|\mathcal{X}|-1$ variables. Let us define $\lambda_i\triangleq p(V=v_i)$ and $\left[p_1^{(i)} p_2^{(i)} \ldots p_{|\mathcal{X}|}^{(i)}\right]\triangleq p(X=x|V=v_i)$. We have $\lambda_i \geq 0$, $ p_j^{(i)} \geq 0$ and $\sum_i \lambda_i =1$, $\sum_j p_j^{(i)} =1$. The following is a restatement of the constrained optimization problem in (\ref{vv}): \begin{align}  \max_{\{\lambda_i\}, \{p_j^{(i)}\}} \ \ &  f\left(\sum_i \lambda_ip_j^{(i)}\right) - \sum_i \lambda_i f_{\mu}(p_j^{(i)}) \nonumber\\ \nonumber  \mbox{s.t.} \ \ \ &  \sum_i \lambda_i =1, \ \sum_j p_j^{(i)} =1 \\ \label{zz} & \lambda_i \geq 0, \  p_j^{(i)} \geq 0 \end{align}

Note that the cyclic shift symmetry assumption on Bob's and Eve's channels yields a significant reduction in the cardinalities of the auxiliary random variables. In particular, the bound on the rate splitting variable reduces from $|\mathcal{X}|+3$ to $|\mathcal{X}|$ and the bound on the channel prefixing variable reduces from $|\mathcal{X}|^2 + 4|\mathcal{X}| + 3$ to $|\mathcal{X}|^2$. In fact, the problem in (\ref{vv}) for $\mu=0$ is equivalent to finding the secrecy capacity $C_s$. Thus, in cyclic shift symmetric wiretap channels, solving a problem of the same number of variables as finding the secrecy capacity is sufficient to characterize the optimal selections of $U$ and $V$ for any point on the boundary of the rate-equivocation region. Another remark is that the constrained optimization problem in (\ref{zz}) for $\mu=0$ is equivalent to finding the secrecy capacity for general wiretap channels not necessarily cyclic shift symmetric. 

The structure of the optimal auxiliary selections $U^*$ and $V^*$ for cyclic shift symmetric wiretap channels in Theorem \ref{cyc} indicates a sufficient condition for $U=\phi$ to be an optimal selection: If the optimizing $p(\hat{V}=v_k)$ and $p(X=x|\hat{V}=v_k)$, $k=1,\ldots,|\mathcal{X}|$ in (\ref{vv}) are such that \begin{align} \label{str} p(\hat{V}=v_k)=\frac{1}{|\mathcal{X}|}, \quad \mbox{and} \quad p(X=x|\hat{V}=v_k)=p(X=x|\hat{V}=v_1)(k-1), \quad \forall k \end{align} then rate splitting is not necessary. In this case, as $\hat{V}$ has cardinality $|\mathcal{X}|$, each element of $\hat{V}$ generates a cyclic shift of $p(X|\hat{V}=v_1)$, and a uniform distribution is generated over $X$. Therefore, the uniform $\hat{V}$ together with the cyclic prefix channel $\hat{V} \rightarrow X$ in (\ref{str}) maximize $I(X;Y)$ and hence the objective function in (\ref{ree}) (c.f. the proof of Theorem \ref{cyc} in Appendix \ref{app3}). In other words, if (\ref{str}) is satisfied, then $U^*$ and $V^*$ as selected in (\ref{dd1})-(\ref{dd2}) yield a uniform PMF for $p(X|U^*=u)$ for all $u \in \{1,\ldots,|\mathcal{X}|\}$, i.e., $U^*$ is independent of $X$. Therefore, if (\ref{str}) is satisfied, $U=\phi$ can be selected without losing optimality, i.e., $U$ is not necessary.

Next, we consider a sub-class of cyclic shift symmetric channels, namely dominantly cyclic shift symmetric channels (c.f. \cite[Definition 5]{nair10}):
\begin{definition}
A cyclic shift symmetric wiretap channel is dominantly cyclic shift symmetric if $f(\uv) \geq f(P_x)$, $\forall P_x \in \Delta$, where $\uv$ is the $|\mathcal{X}|$ dimensional uniform distribution.
\end{definition}
Note that from \cite[Theorem 3]{vandijk97} and the fact that the uniform distribution is capacity achieving for cyclic shift symmetric channels, a less noisy cyclic shift symmetric wiretap channel is also dominantly cyclic shift symmetric (see also \cite{spec}). We denote dominantly cyclic shift symmetric channels by the shaded region in Fig. \ref{sch}. Note that all cyclic shift symmetric channels in regions \textcircled{1} and \textcircled{2} are shaded. In the following lemma, we prove the sufficiency of dominant cyclic shift symmetry for having the solution of (\ref{vv}) satisfy the property in (\ref{str}). We provide the proof of this lemma in Appendix \ref{app4}.
\begin{lemma} \label{ism}
For dominantly cyclic shift symmetric wiretap channels, a solution of the problem in (\ref{vv}) is a distribution $p(\hat{V})$ and a prefix channel $\hat{V} \rightarrow X$ that satisfy the condition in (\ref{str}). 
\end{lemma}

A corollary of Lemma \ref{ism} and the structure in Theorem \ref{cyc} is that rate splitting is unnecessary for dominantly cyclic shift symmetric channels and the entire rate-equivocation region can be attained by channel prefixing alone.
\begin{corollary} \label{split}
In a dominantly cyclic shift symmetric wiretap channel,
rate splitting does not improve the rate-equivocation region and optimal channel prefixing has the cardinality $|\mathcal{V^*}| \leq |\mathcal{X}|$.
In particular,
\begin{align}
\label{css} C_s = \max_{P_x} f(P_x) - \min_{P_x} f(P_x)
\end{align}
\end{corollary}

We remark here that if the wiretap channel is dominantly cyclic symmetric, then known inner and outer bounds on the corresponding broadcast channel capacity region are shown to coincide in \cite{nair10}. Therefore, the broadcast channel capacity region, which is in general an open problem, can be fully characterized for dominantly cyclic shift symmetric channels. We observe here that dominant cyclic symmetry yields a similar simplification for the wiretap channel, rendering rate splitting variable $U$ unnecessary. However, note that the class of cyclic shift symmetric wiretap channels for which rate splitting is unnecessary is strictly larger than the class of dominantly cyclic shift symmetric channels. In fact, for all cyclic shift symmetric channels which satisfy (\ref{str}), $U=\phi$ is optimal and dominant cyclic shift symmetry is just a sufficient but not necessary condition for the property (\ref{str}). In Section \ref{bin}, we provide examples for binary-input cyclic shift symmetric wiretap channels that are not dominantly cyclic shift symmetric but for which rate splitting is still unnecessary.

Note that the secrecy capacity expression in (\ref{css}) is the solution of the problem in (\ref{vv}) for $\mu=0$. We also remark that (\ref{css}) is generally an upper bound for a wiretap channel:
\begin{align}
C_s & = \max ~I(V;Y) - I(V;Z) \\
 & = \max ~[I(X;Y)-I(X;Z)] - [I(X;Y|V)-I(X;Z|V)] \\
 & \leq \max_{P_x}f(P_x) - \min_{P_x}f(P_x) \label{upb}
\end{align}
but it is attained for dominantly cyclic shift symmetric wiretap channels by Corollary \ref{split}.

Next, we consider more capable cyclic shift symmetric channels. $V^*=X$ is optimal in this case due to Theorem \ref{tm_mc}; hence, the structure in Theorem \ref{cyc} for general cyclic shift symmetric channels is further reduced under the more capable condition. In addition, the most componentwise dominant rate-equivocation pair $(C_B,C_s)$ is achieved in more capable cyclic shift symmetric channels as we show in the following corollary. Further, if the channel is more capable and dominantly cyclic shift symmetric, $(C_B,C_s)$ pair is achieved by $V^*=X$ and $U^*=\phi$. We provide the proof of this corollary in Appendix \ref{appd}.
\begin{corollary} \label{th_mc}
In a more capable cyclic shift
symmetric wiretap channel, $V^*=X$ and the rate-equivocation pair $(C_B,C_s)$ is
achievable. In a more capable dominantly cyclic shift symmetric wiretap channel, $(C_B,C_s)$ pair is achieved by $V^* = X$ and $U^* = \phi$. 
\end{corollary}

We remark here that Theorem \ref{cyc} applies to more capable cyclic shift symmetric wiretap channels as well. Even though $V^*=X$ is stated as optimal in Corollary \ref{th_mc}, the general structure of $U^*$ and $V^*$ in Theorem \ref{cyc} does not reduce to $V^* = X$ when more capable condition is imposed. Note that the cardinality of $V^*$ in Theorem \ref{cyc} is $|\mathcal{X}|^2$ while $V^* = X$ requires only a cardinality of $|\mathcal{X}|$ for $V^*$. However, $(C_B,C_s)$ pair is achieved with both auxiliary selections. Therefore, $U^*$ and $V^*$ in Theorem \ref{cyc} are not necessarily unique optimal auxiliary selections. 
 
More capable cyclic shift symmetric wiretap channels
have already been covered in \cite{vandijk97} in the following
example:
\[p(y|x)=\frac{1}{2} \left( \begin{array}{cccc}
1-p & p & 1-q & q \\
p & 1-p & q & 1-q \\
1-q & q & 1-p & p \\
q & 1-q & p & 1-p \end{array} \right)\]
and
\[p(z|x)=\frac{1}{2} \left( \begin{array}{cccc}
1-r & 1-r & r & r \\
1-r & 1-r & r & r \\
r & r & 1-r & 1-r \\
r & r & 1-r & 1-r \end{array} \right)\] In \cite{vandijk97}, it
is shown that, for $r$ close enough to $1/2$ (depending on the
values of $p$ and $q$), the wiretap channel is more capable.
However, in the same reference, the channel is shown to be not less noisy for any $r$,
$p$ and $q$. We now observe that $p(y|x)$ and $p(z|x)$ are
cyclic shift symmetric channels. Therefore, by
Corollary~\ref{th_mc}, $(C_B,C_s)$ pair is achievable by a
non-trivial $U^*$ with uniform distribution and $V^*=X$ when $r$, $p$ and $q$ are such that
the wiretap channel is more capable.

We note that the results obtained for discrete alphabet cyclic shift symmetric channels naturally extend if the alphabets are bounded continuous intervals. In particular, the definition of cyclic shift symmetry extends naturally for $\mathcal{X}=[0,b)$: If $I(X;Y)$ is invariant under any modular shift in the input PDF, the channel is cyclic shift symmetric. Typical examples of continuous alphabet cyclic shift symmetric channels are modulo additive noise channels \cite{erezzamir00}. If cyclic shift symmetry holds, the channel capacity is achieved at uniform distribution over $\mathcal{X}$. Hence, if both the main and eavesdropping channels are cyclic shift symmetric, then the optimal selections $U^*$ and $V^*$ have the same structure as in Theorem \ref{cyc}. The definition of dominant cyclic shift symmetry also extends similarly for continuous alphabets and rate splitting is not necessary for continuous alphabet dominantly cyclic shift symmetric wiretap channels. 

The result does not directly extend for unbounded input alphabets, i.e., for $b=\infty$, with an average power constraint. Even if the cyclic shift symmetry holds, it may not be possible to generate Bob's capacity achieving input PDF by shifting the solution of the auxiliary optimization problem and therefore the proof method in Theorem \ref{cyc} is not directly applicable.

\section{Binary-Input Cyclic Shift Symmetric Wiretap Channels} \label{bin}

\begin{figure}[t]
\centering
\includegraphics[width=0.65\linewidth]{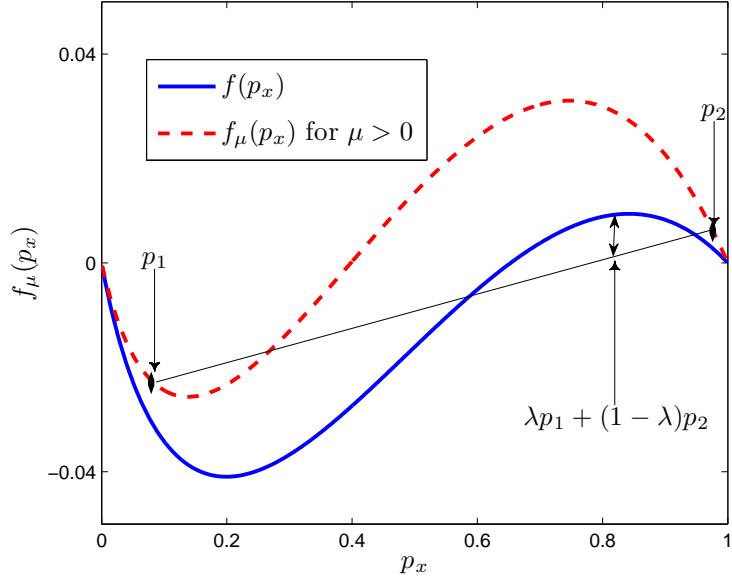}
\caption{Visualization of the optimization problem in terms of $p_1$, $p_2$ and $\lambda$.}
\label{vis}
\end{figure}

In this section, we consider cyclic shift symmetric wiretap channels with binary input: $|\mathcal{X}|=2$. Note that the cardinality requirement on $V$ to solve the problem in (\ref{vv}) is $|\mathcal{V}|=2$ for binary input wiretap channels. Let $p(v_1)=\lambda$, $p(x|v_1)=[p_1, 1-p_1]$ and $p(x|v_2)=[p_2, 1-p_2]$. Let the resulting input distribution be $P_x=[p_x, 1-p_x]$. The optimization problem in (\ref{vv}) and (\ref{zz}) for the binary-input case reduces to: 
\begin{align} \nonumber
 \max_{\lambda,p_1,p_2} & \ f(\lambda p_1 + (1-\lambda)p_2) - \lambda f_{\mu}(p_1) - (1-\lambda) f_{\mu}(p_2) \\
\mbox{s.t.} & \ \ 0 \leq \lambda, p_1,p_2 \leq 1\label{eqopt}
\end{align}
A geometrical visualization for the problem in (\ref{eqopt}) is provided in Fig. \ref{vis}. Two points are picked from the $x$-axis and their image on $f_{\mu}$ are combined to form a line. $\lambda$ determines the point of operation and the value of the objective function is the difference between $f$ and the formed line segment at that particular point. 

\begin{figure}[t]
\centering
\includegraphics[width=0.68\linewidth]{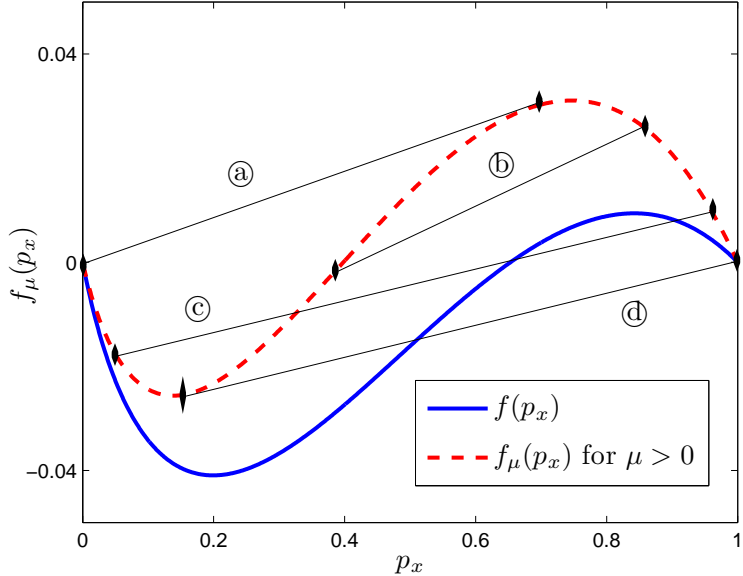}
\caption{Optimality conditions for $p_1$, $p_2$ and $\lambda$. \textcircled{a} and \textcircled{d} satisfy the optimality conditions while \textcircled{b} and \textcircled{c} do not satisfy the optimality condition.}
\label{vis2}
\end{figure}

The necessary optimality conditions for the problem in (\ref{eqopt}) are found by taking the derivative of the objective function with respect to $p_1, p_2$ and $\lambda$, respectively. If $ p_1^*, p_2^*, \lambda^*$ are strictly interior to the $[0,1]$ interval, i.e., not equal to 0 or 1, then \begin{align} \lambda^*\left(f'(\lambda^*p_1^* + (1-\lambda^*)p_2^*) - f'_{\mu}(p_1^*)\right)&=0 \\ (1-\lambda^*)\left(f'(\lambda^*p_1^* + (1-\lambda^*)p_2^*) - f'_{\mu}(p_2^*)\right) &=0 \\ (p_1^* - p_2^*)f'(\lambda^*p_1^* + (1-\lambda^*)p_2^*) - (f_{\mu}(p_1^*) - f_{\mu}(p_2^*)) &= 0 \label{fin} \end{align}
Note that $\lambda^* \neq 0,1$ as the objective function in (\ref{eqopt}) takes the value zero for $\lambda=0,1$. Hence, the optimality condition in (\ref{fin}) always holds and we get:  \begin{align} \label{lw1} f'(\lambda^*p_1^* + (1-\lambda^*)p_2^*) = \frac{f_{\mu}(p_1^*) - f_{\mu}(p_2^*)}{p_1^* - p_2^*}  \end{align} If, in addition, $p_1^*, p_2^* \neq 0,1$, \begin{align} f'(\lambda^*p_1^* + (1-\lambda^*)p_2^*)= f'_{\mu}(p_1^*) = f'_{\mu}(p_2^*) \label{lw2} \end{align}If $p_1^* = 0,1$ and $p_2^*\neq 0,1$, then \begin{align} f'(\lambda^*p_1^* + (1-\lambda^*)p_2^*)= f'_{\mu}(p_2^*) \end{align} and similarly if $p_2^* =0,1$ and $p_1^* \neq 0,1$, then \begin{align} f'(\lambda^*p_1^* + (1-\lambda^*)p_2^*)= f'_{\mu}(p_1^*) \label{lw3} \end{align} The conditions in (\ref{lw1})-(\ref{lw3}) have the following geometric interpretation: In Fig. \ref{vis}, the line drawn from $(p_1^*,f_{\mu}(p_1^*))$ and $(p_2^*,f_{\mu}(p_2^*))$ must be tangent to the $f_{\mu}$ curve at both points. If $p_1^*$ or $p_2^*$ are $0$ or $1$, then this tangency does not have to hold at that point. We illustrate these conditions in Fig. \ref{vis2}. We observe that for the selections of $p_1$ and $p_2$ in the configurations \textcircled{a} and \textcircled{d}, the line is tangent to $f_{\mu}(p_x)$ at only one point and the other point is either 0 or 1. However, \textcircled{b} and \textcircled{c} do not satisfy the optimality condition as the $p_1$ and $p_2$ points lie interior to $[0,1]$ but the line is not tangent to $f_{\mu}$. In fact, we observe by inspection that \textcircled{a} and \textcircled{d} are the only possible configurations in the particular values chosen in Fig. \ref{vis2} that satisfy the optimality conditions in (\ref{lw1})-(\ref{lw3}).

Note that the geometric interpretation of the optimality conditions provide a simple check if a point $p \in (0,1)$ is one of the optimal selections $p_1^*,p_2^*$: Draw the tangent line for $f_{\mu}$ at $p$. If this tangent line does not intersect $f_{\mu}$ other than $p$ or if it intersects at a point $p'\in (0,1)$ but it is not tangent at $p'$, then $p$ cannot be an optimal selection. Also note that optimality conditions do not rule out the trivial selection $p_1=0$ and $p_2=1$. Hence, this selection is always a candidate to be an optimal selection. This selection is indeed optimal if $f(p_x)\geq 0$ for all $p_x \in [0,1]$, i.e., when the wiretap channel is more capable by Theorem \ref{tm_mc}. Geometrically, when $f(p_x) \geq 0$ the points $(p_i,f_{\mu}(p_i))$ have nonnegative $y$-axis as $f_{\mu}(p_x)\geq f(p_x) \geq 0$. The level of the line segment is the smallest when $p_1=0$, $p_2=1$. Moreover, the points $\lambda p_1 + (1-\lambda)p_2$ span the space of two-dimensional PMFs when $p_1=0$ and $p_2=1$. Hence, the difference of the function and the line segment has the highest value when $p_1=0$ and $p_2=1$, that is, the optimal selection is $p_1=0$ and $p_2=1$.

\subsection{The BSC-BEC Wiretap Channel}

Let the main channel be BSC($\epsilon$) and the eavesdropper's channel be BEC($\alpha$). Note that both BSC and BEC are cyclic shift symmetric. $\mathcal{X}=\mathcal{Y}=\{0,1\}$ and $\mathcal{Z}=\{0,e,1\}$.  
For $0 \leq \epsilon < 0.5$ and the
input distribution $P_x=[p_x,1-p_x]$, we have
\begin{align}\label{fcn}
\hspace*{-0.1cm} f(p_x)= h((2\epsilon-1)p_x + 1 - \epsilon) - h(\epsilon) - (1-\alpha)h(p_x) 
\end{align}
where $h(.)$ is the binary entropy function. We first
investigate some geometric properties of the function $f(p_x)$
in (\ref{fcn}). It can be shown \cite{nair10} that when $p(y|x)$ is BSC($\epsilon$) and $p(z|x)$ is BEC($\alpha$):
\begin{enumerate}
    \item If $\alpha<4\epsilon(1-\epsilon)$, then Eve is less noisy than Bob.
    \item If $4\epsilon(1-\epsilon) \leq \alpha \leq h(\epsilon)$, Eve is more capable but not less noisy than Bob.
    \item If $h(\epsilon) < \alpha$, the wiretap channel is dominantly cyclic shift symmetric.
\end{enumerate}

\begin{figure}[t]
\centering
\includegraphics[width=0.69\linewidth]{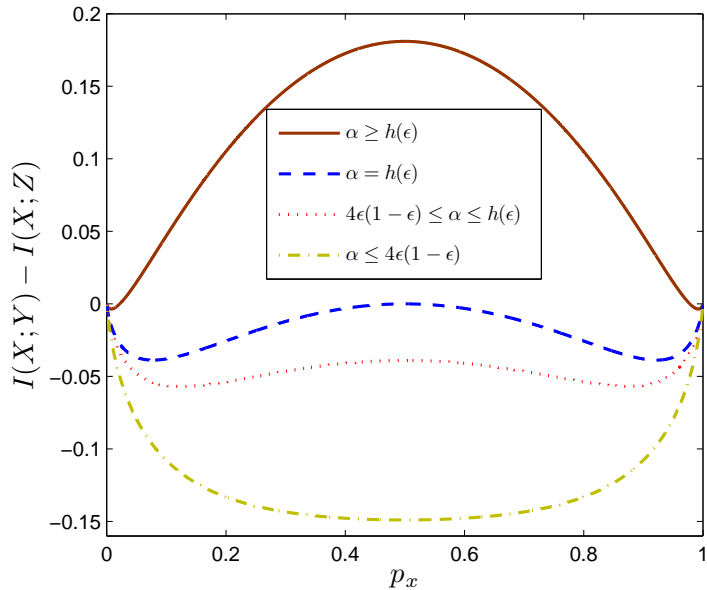}
\caption{$f(p_x)$ as a function of $p_x$ for the BSC-BEC channel.}
\label{figfig}
\end{figure}

An illustration of $f(p_x)$ function for the BSC-BEC channel is provided in
Fig.~\ref{figfig} for $\epsilon=0.1$ and various $\alpha$. Note
that for any $\epsilon$ and $\alpha$, $f(p_x)<0$ for some
$p_x$; thus, the channel is not more capable and is always in
region \textcircled{4} in Fig.~\ref{sch}. We observe that for
$\alpha>h(\epsilon)$, $f(p_x)$ is maximized at $p_x=0.5$, i.e., the channel satisfies dominant cyclic shift symmetry. From Corollary~\ref{split}, rate splitting is not necessary for
$\alpha>h(\epsilon)$ and moreover the required channel prefixing has $|\mathcal{V^*}|=2$ with $p(v_1)=p(v_2)=1/2$ and $p(x|v_1)=[a,1-a]$, $p(x|v_2)=[1-a,a]$ where $[a,1-a]$ is an input distribution that maximizes $\left[(\mu + 1)I(X;Y) - I(X;Z)\right]$. The secrecy capacity is
\begin{align}
\label{ccc}
C_s = \max_{p_x} f(p_x) - \min_{p_x} f(p_x)
\end{align}
We also note that for $\alpha<4\epsilon(1-\epsilon)$, Eve is less noisy than Bob, and the secrecy capacity is $C_s=0$ \cite{CK78}. We investigate the remaining case, which is $4\epsilon(1-\epsilon) \leq \alpha \leq h(\epsilon)$, in the next subsection.

\begin{figure}[h]
\centering
\includegraphics[width=0.65\linewidth]{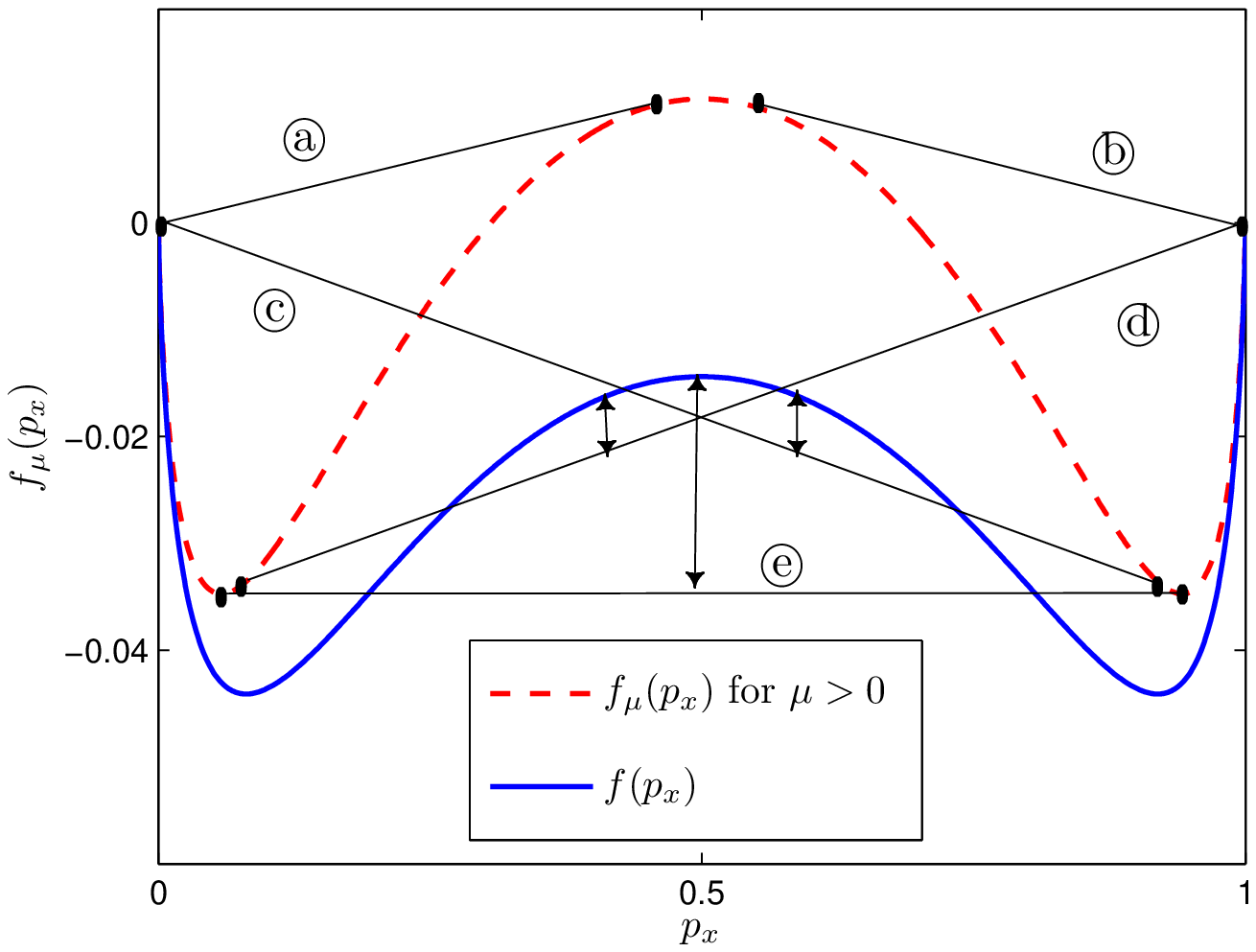}
\caption{Five configurations that satisfy optimality conditions. \textcircled{e} is always the optimal.}
\label{conf}
\end{figure}

\subsubsection{The Case of $4\epsilon(1-\epsilon) \leq \alpha \leq h(\epsilon)$}

When $4\epsilon(1-\epsilon) \leq \alpha \leq h(\epsilon)$ in the BSC-BEC channel, neither Eve is less noisy nor the dominant cyclic shift symmetry holds. Secrecy capacity is still non-zero in this case and the rate-equivocation region has a non-empty interior. One can verify easily by tracing all points in $[0,1]$ that there are only 5 configurations that satisfy the necessary optimality conditions as well as the trivial selection $V=X$, i.e., $p_1=0$ and $p_2=1$. However, the trivial selection is immediately eliminated as $f_{\mu}(p_x)\leq 0$ for some $p_x$ in this case, and hence $V=X$ is strictly suboptimal by Theorem \ref{tm_mc}. 

The other 5 configurations are shown in Fig. \ref{conf}: In configurations \textcircled{a}, \textcircled{b}, \textcircled{c} and \textcircled{d}, either $p_1$ or $p_2$ is on the boundary and in configuration \textcircled{e}, both $p_1$ and $p_2$ are in the interior with the property $p_1 \in \arg\min_{p_x \in [0,1]} f_{\mu}(p_x)$ and $p_2 = 1-p_1$. By comparing these three configurations, we observe that the optimum selection is always configuration \textcircled{e}. In other words, we have for $\mu \geq 0$ and for all $0\leq \lambda,p_1,p_2 \leq 1$, \begin{align} f(0.5)-\min_{p_x} f_{\mu}(p_x) \geq f(\lambda p_1 + (1-\lambda)p_2) - \lambda f_{\mu}(p_1) - (1-\lambda)f_{\mu}(p_2)  \end{align}   

\begin{figure}[t]
\centering
\includegraphics[width=0.6\linewidth]{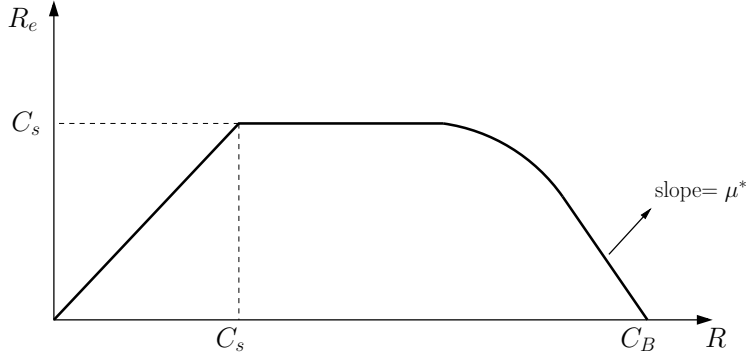}
\caption{The general form of the rate-equivocation region of the BSC-BEC wiretap channel for $4\epsilon(1-\epsilon) \leq \alpha \leq h(\epsilon)$.}
\label{regext}
\end{figure}

Note that \textcircled{e} has the desirable property that $p_2=1-p_1$ and $\lambda^* = \frac{1}{2}$. This property is equivalent to the one in (\ref{str}) in the binary-input case. Therefore, $U=\phi$ is optimal, and the upper right boundary of the rate-equivocation region can be traced by $V$ only. However, unlike the case of $h(\epsilon)\leq \alpha$, if $4\epsilon(1-\epsilon) \leq \alpha \leq h(\epsilon)$, there exists $\mu\geq 0$ such that $f(0.5) < \min_{p_x} f_{\mu}(p_x)$. We define $\mu^*$ as \begin{align} \mu^* = \min\{\mu : f(0.5) \leq \min_{p_x} f_{\mu}(p_x)\} \end{align} For $\mu > \mu^*$, $V$ defined as above cannot improve the objective function. Thus, trivial $V$ is the optimal selection for $\mu > \mu^*$. However, the highest achievable equivocation with a trivial $V$ selection is zero as Eve's channel is more capable with respect to Bob's channel in this case. Hence, for $\mu > \mu^*$, the only possible achievable point is $(C_B,0)$. The general form of the rate-equivocation region is given in Fig. \ref{regext}. The upper right boundary includes the line segment that combines the point for which the supporting line slope is $\mu^*$ and the $(C_B,0)$ point. This line segment has the slope $\mu^*$.      

In conclusion, rate splitting $U$ is not necessary for determining the rate-equivocation region of the BSC-BEC wiretap channel and in particular the secrecy capacity is 
\begin{align} C_s = f(0.5) - \min_{p_x} f(p_x) \label{agr} \end{align}
Note that (\ref{agr}) is in agreement with (\ref{ccc}), as in that case $\max_{p_x} f(p_x)$ is achieved at $p_x = 0.5$. 

\subsection{The BEC-BSC Wiretap Channel}

Now, let the main channel be BEC($\alpha$) and the
eavesdropper's channel be BSC($\epsilon$).
$\mathcal{X}=\mathcal{Z}=\{0,1\}$ and $\mathcal{Y}=\{0,e,1\}$. We have the following facts \cite{nair10}:
\begin{enumerate}
    \item If $\alpha<4\epsilon(1-\epsilon)$, then Bob is less noisy than Eve.
    \item If $4\epsilon(1-\epsilon) \leq \alpha \leq h(\epsilon)$, then Bob is more capable but not less noisy than Eve. 
\end{enumerate}

Hence, if $4\epsilon(1-\epsilon) \leq
\alpha \leq h(\epsilon)$, the wiretap channel is in region
\textcircled{3} in Fig.~\ref{sch}.
By \cite{CK78}, $C_s = \max_{P_x} f(p_x)$, and from
Corollary~\ref{th_mc}, $(C_B,C_s)$ is achievable. If
$\alpha<4\epsilon(1-\epsilon)$, as both channels are cyclic
shift symmetric, by \cite[Theorem 3]{vandijk97}, $C_s = C_B -
C_E$ and $(C_B,C_s)$ is achievable. We investigate the remaining case, which is $\alpha \geq h(\epsilon)$, in the next subsection.

\subsubsection{The Case of $\alpha \geq h(\epsilon)$}

\begin{figure}[t]
\centering
\includegraphics[width=0.6\linewidth]{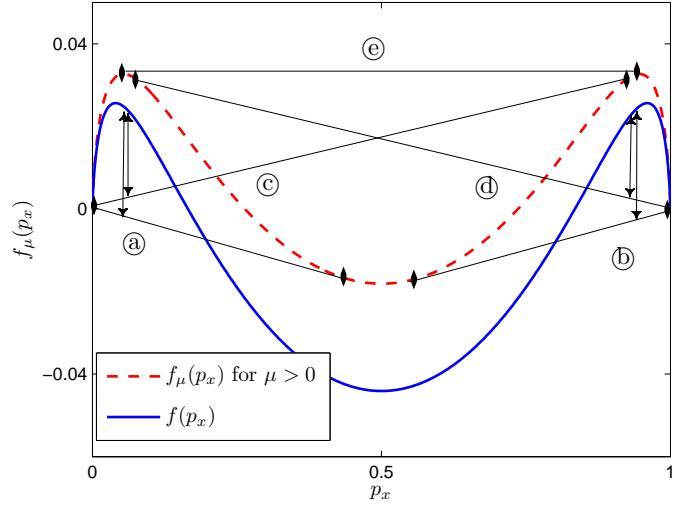}
\caption{Five configurations that satisfy the necessary optimality conditions for the BEC-BSC wiretap channel. \textcircled{a} and \textcircled{b} are both optimal.}
\label{confg2}
\end{figure}

In the BEC-BSC wiretap channel, if $\alpha \geq h(\epsilon)$, neither less noisy nor more capable condition holds. We first solve the optimization problem in (\ref{eqopt}) by inspecting the tangent lines drawn at interior points $p \in (0,1)$. One can easily verify that, as in the BSC-BEC channel, there are only 5 possible configurations that satisfy the necessary optimality conditions in addition to the trivial selection $p_1=0$ and $p_2=1$. As Bob's channel is not more capable with respect to Eve in this case and $f(p_x)$ is maximized at an interior point, by Theorem \ref{tm_mc}, trivial selection is strictly suboptimal in this case. In particular, the trivial selection is not optimal for all $\mu$ such that $f_{\mu}(p_1)<0$ for some $p_1 \in [0,1]$. We show the other 5 configurations in Fig. \ref{confg2}. In configuration \textcircled{e}, $p_1, p_2 \in (0,1)$ with $f_{\mu}(p_1)=f_{\mu}(p_2)$ and $f'_{\mu}(p_1)=f'_{\mu}(p_2)=0$. Hence, it satisfies the optimality condition for $\lambda=0.5$. However, the objective function $f(0.5) - 0.5\left(f_{\mu}(p_1) + f_{\mu}(p_2)\right)<0$; therefore, this configuration cannot be optimal. The other configurations \textcircled{a}, \textcircled{b}, \textcircled{c} and \textcircled{d} have $p_1$ or $p_2$ on the boundary of $[0,1]$ interval as shown in Fig. \ref{confg2}. We observe that \textcircled{a} and \textcircled{b} achieve the same value of the objective function and it is always higher compared to that achieved by \textcircled{c} and \textcircled{d}. Therefore, \textcircled{a} and \textcircled{b} are optimal selections for the problem in (\ref{eqopt}). Note that \textcircled{a} is obtained by cyclic shifts of \textcircled{b}.   
The configuration in \textcircled{a} is also represented as $p(X=0|V=v_1)=0$ and $p(X=0|V=v_2)=p_1$ where the line segment that combines $(0,0)$ and $(p_1,f_{\mu}(p_1))$ is tangent to the curve $(p_x,f_{\mu}(p_x))$. Similarly, \textcircled{b} is equivalent to $p(X=0|V=v_1)=1$ and $p(X=0|V=v_2)=1-p_1$. The rate equivocation region is traced by varying $\mu$ and finding $p_1$ that satisfies the tangency and $\lambda^*$ that yields the optimal value of the objective function given $p_1$. In particular, we define $\mu^*$ as \begin{align} \mu^* = \min\{\mu \geq 0 | \min_{p_x}f_{\mu}(p_x) \geq 0\} \end{align} 

For $\mu\leq \mu^*$, we use the following $U$ and $V$: \begin{align} p(U=u_1)&=p(U=u_2)=0.5 \\ p(V=v_1|U=u_1)&=\lambda^*, \quad p(V=v_2|U=u_1)=1-\lambda^*, \\  p(V=v_3|U=u_2)&=\lambda^*, \quad p(V=v_4|U=u_2)=1-\lambda^*, \\ p(X=0|V=v_1)&=0, \quad p(X=0|V=v_2)=p_1, \\ p(X=0|V=v_3)&=1, \quad p(X=|V=v_4)=1-p_1 \end{align}

\begin{figure}[t]
\centering
\includegraphics[width=0.6\linewidth]{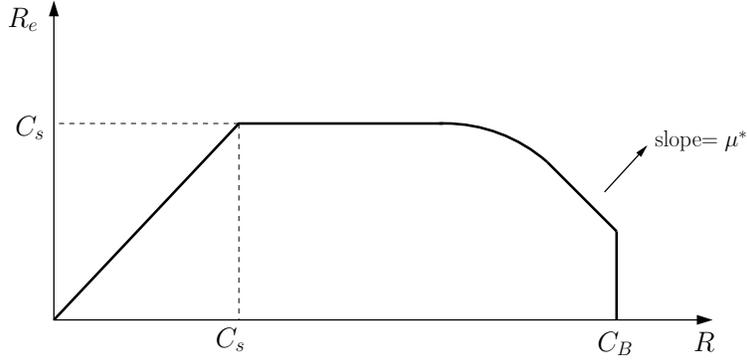}
\caption{General form of the rate-equivocation region of the BEC-BSC wiretap channel when $\alpha \geq h(\epsilon)$.}
\label{genf}
\end{figure}

For $\mu>\mu^*$, $V$ is not necessary as $f_{\mu}(p_x)\geq 0$ in this case. We obtain a case similar to the more capable condition and one can easily show that a non-trivial $V$ does not improve the objective function. As $V$ is not used for $\mu>\mu^*$, the achieved rate $I(V;Y)=I(X;Y)$ and optimal selection of $U$ as in Theorem \ref{cyc} generates uniform distribution on the channel input $X$, which is capacity achieving for Bob's channel. Hence, for $\mu>\mu^*$, $C_B$ is achieved. The general form of the rate-equivocation region is depicted in Fig. \ref{genf}. Note that the supporting line with slope $\mu^*$ is on the boundary of the rate-equivocation region. 

\subsection{The Existence of More Capable but Not Less Noisy Dominantly Cyclic Shift Symmetric Wiretap Channels}

For the BSC-BEC and BEC-BSC wiretap channels, we observe that if the channel is more capable but not less noisy, dominant cyclic symmetry does not hold. Conversely, if dominant cyclic symmetry holds the channel is not more capable. We question whether this property extends for general cyclic shift symmetric wiretap channels, i.e., we ask whether there exist more capable not less noisy cyclic shift symmetric wiretap channels that satisfy dominant cyclic shift symmetry. In this subsection, we answer this question in the affirmative direction. 

We consider the following class of binary-input cyclic shift symmetric wiretap channels: 
\[p(y|x)= \left( \begin{array}{cc}
1-\epsilon & \epsilon \\
\epsilon & 1-\epsilon \\
 \end{array} \right)\]
and
\[p(z|x)= \left( \begin{array}{ccc}
1-p-q & q & p \\
q & 1-p-q & p  \end{array} \right)\]
Bob's channel $p(y|x)$ is BSC($\epsilon$). Eve's channel $p(z|x)$ is parameterized by $p$ and $q$ and if $p=0$, it reduces to BSC($q$). Using the same notation $P_x=[p_x,1-p_x]$ and $f(p_x)=I(X;Y) - I(X;Z)$, we have the first and second derivatives of $f(p_x)$ as \begin{align} \nonumber \frac{d}{dp_x}f(p_x) = (1&-2\epsilon)\log\left( \frac{(2p_x-1)\epsilon + 1-p_x}{p_x + (1-2p_x)\epsilon} \right) \\ & - (1-p-2q)\log\left(\frac{p_x (p+2q-1) + (1-p-q)}{p_x(1-p-2q) + q}\right) \\ \nonumber \frac{d^2}{d p_x^2} f(p_x) = -&\frac{(2\epsilon -1)^2}{\left((2p_x -1)\epsilon + 1-p_x\right)\left(p_x + (1-2p_x)\epsilon\right)} \\ &+ \frac{(1-p-2q)^2}{\left(p_x (p+2q-1) + (1-p-q)\right)\left(p_x(1-p-2q) + q\right)} \label{sder}\end{align}
Since both $p(y|x)$ and $p(z|x)$ are cyclic shift symmetric, we have $\frac{d}{dp_x}f(p_x)=0$ at $p_x=0.5$. In addition, as the second derivative in (\ref{sder}) is quadratic in $p_x$, it can vanish at at most two points in the $[0,1]$ interval. In fact, due to the cyclic symmetry, the second derivative vanishes at $p_x^*$ and $1-p_x^*$ or it does not vanish. Hence, $f(p_x)$ can have at most one inflection point in the $(0,0.5)$ interval. 

\begin{figure}[t]
\centering
\includegraphics[width=0.6\linewidth]{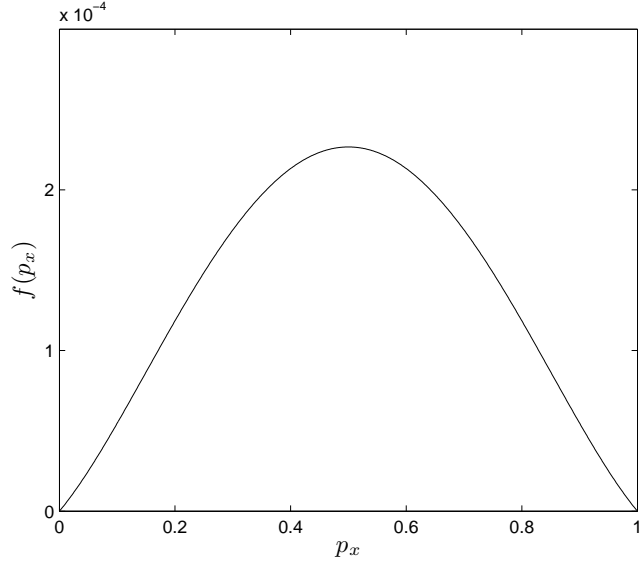}
\caption{$f(p_x)$ for a more capable but not less noisy dominantly symmetric wiretap channel.}
\label{cntr}
\end{figure}

If $f(p_x)$ has strictly positive first and second derivatives at $p_x=0$, then it has one inflection point in the $(0,0.5)$ interval. This is due to the fact that $\frac{d}{d p_x} f(p_x)>0$  at $p_x=0$ while $\frac{d}{d p_x} f(p_x) = 0$ at $p_x = 0.5$. As $\frac{d}{d p_x} f(p_x)$ is increasing at $p_x=0$, $\frac{d}{d p_x} f(p_x)$ should finish its ascent and start its descent in the interval $(0,0.5)$. As the second derivative is a continuous function of $p_x$ and due to the mean value theorem, it vanishes at a point in the $(0,0.5)$ interval. As the second derivative may vanish at at most one point in the $(0,0.5)$ interval, $\frac{d}{d p_x} f(p_x) >0$ for $p_x \in (0,0.5)$ and $f(p_x)$ is maximized at $p_x=0.5$. Hence, dominant cyclic shift symmetry holds if $f(p_x)$ has strictly positive first and second derivatives at $p_x=0$. We next show that there exist parameters $\epsilon$, $p$ and $q$ such that the first and second derivatives of $f(p_x)$ at $p_x=0$ are strictly positive. The derivatives at $p_x=0$ are expressed as  
\begin{align} \frac{d}{dp_x}f(p_x) \big|_{p_x=0} = (1&-2\epsilon)\log\left( \frac{1-\epsilon}{\epsilon} \right) - (1-p-2q)\log\left(\frac{1-p-q}{q}\right) \\ \frac{d^2}{d p_x^2} f(p_x)\big|_{p_x=0} = -& \frac{(2\epsilon -1)^2}{\epsilon(1-\epsilon)} + \frac{(1-p-2q)^2(1-p)}{(1-p-q)q} \label{sevatz} \end{align}
Note that $\frac{d}{dp_x}f(p_x) \big|_{p_x=0}$ is strictly convex monotone decreasing and $\frac{d^2}{d p_x^2} f(p_x)\big|_{p_x=0}$ is strictly concave monotone increasing with $\epsilon \in (0,0.5)$ for given $p,q \in (0,1)$ with $p+q<1$. Moreover, we have \begin{align}  0 & \leq (1-p-2q)\log\left(\frac{1-p-q}{q}\right) \label{sth0} \\ \label{sth1} & \leq \frac{(1-p-2q)^2}{q}\\ \label{sth2} & \leq \frac{(1-p-2q)^2(1-p)}{(1-p-q)q} \end{align} where (\ref{sth0}) is due to the fact that $x\log(1+x) \geq 0$ for $x \geq -1$, (\ref{sth1}) is due to the inequality $\log(1+x) \leq x$ for $x>0$ and (\ref{sth2}) is due to $\frac{1-p}{1-p-q}>1$. Let $c(p,q)=\frac{(1-p-2q)^2(1-p)}{(1-p-q)q}$. Given $p$ and $q$, the second derivative evaluated at $p_x=0$ in (\ref{sevatz}) vanishes at $\epsilon^* \in (0,0.5)$ \begin{align} \epsilon^*(p,q) = \frac{1}{2} - \frac{1}{2}\sqrt{\frac{c(p,q)}{4+c(p,q)}} \end{align} When $\epsilon$ is selected as $\epsilon^*(p,q)$, the first derivative $\frac{d}{dp_x}f(p_x) \big|_{p_x=0}$ is \begin{align}\label{ghg} b(p,q)=(1-2\epsilon^*(p,q))\log\left(\frac{1-\epsilon^*(p,q)}{\epsilon^*(p,q)}\right) - (1-p-2q)\log\left(\frac{1-p-q}{q}\right) \end{align}
One can show that $b(p,q)$ is strictly convex individually in $p$ and $q$ and its first derivatives with respect to $p$ and $q$ vanish at $p+2q=1$. As $b(p,q)=0$ when $p+2q=1$ and $b(p,q) \rightarrow \infty$ for both $p+q=1$ and $p+q=0$, we have $b(p,q) >0$ for $p+2q \neq 1$. Therefore, for given $p$ and $q$ with $p+2q \neq 1$, if $\epsilon$ is chosen as $\epsilon^*(p,q)$, $f(p_x)$ has positive first derivative at $p_x=0$ and second derivative at $p_x=0$ as zero. Since the second derivative is continuous monotone increasing and the first derivative is continuous monotone decreasing in $\epsilon$ given $p$ and $q$, there exists $\delta>0$ such that if $\epsilon$ is chosen as $\epsilon^*(p,q) + \delta$, both derivatives are positive. Therefore, for given $p$ and $q$ with $p+2q \neq 1$, there exists an interval $S = (\epsilon^*(p,q),\epsilon^*(p,q) + \delta) \subset [0,0.5]$ such that if $\epsilon \in S$, the resulting wiretap channel is more capable, not less noisy and dominant cyclic shift symmetric. 
We illustrate $f(p_x)$ for such a wiretap channel in Fig. \ref{cntr}. The parameters are $\epsilon=0.4202$, $p=0.6$ and $q=0.25$ for this figure. For $p=0.6$ and $q=0.25$, $\epsilon^*(p,q)=0.4194$; hence $\delta > 0.0008$ in this case. Note from Fig. \ref{cntr} that the first and second derivatives of $f(p_x)$ are positive at $p_x=0$ as it is monotone increasing and convex in a neighborhood of $p_x=0$. Then, the inflection point is observed and the function becomes concave driving the first derivative to zero at $p_x=0.5$.  

In conclusion, we show that the intersection of the shaded region and region \textcircled{3} in Fig. \ref{sch} is non-empty and for these channels neither rate splitting nor channel prefixing is necessary\footnote{In a conference version of this work \cite{Omur11ISIT}, we mistakenly claimed that rate splitting is strictly necessary for more capable but not less noisy channels. This also disproves \cite[Corollary 1]{Omur11ISIT}.}.

%\begin{figure}[t]
%\centering
%\includegraphics[width=0.6\linewidth]{BEC_BSC_diff_mu.eps}
%\caption{ BEC(0.18)-BSC(0.67) wiretap channel.}
%\label{samp5}
%\end{figure}
%
%\begin{figure}[t]
%\centering
%\includegraphics[width=0.6\linewidth]{BEC_BSC_RE_example.eps}
%\caption{Rate-equivocation region of BSC(0.18)-BEC(0.67) wiretap channel.}
%\label{samp6}
%\end{figure}

\section{Conclusions}

In this paper, we provided new results on the computation of optimal rate
splitting and channel prefixing auxiliary random variables in evaluating the boundary
of the rate-equivocation region of the discrete memoryless wiretap channel. We proved that
if the wiretap channel is more capable, then channel prefixing is
unnecessary and the entire boundary of the rate-equivocation region is traced by rate splitting alone. Conversely, if the channel is not more capable, we proved under a mild condition that a non-trivial channel prefixing is strictly necessary. Next, we focused on cyclic shift symmetric wiretap channels and explicitly determined the optimal rate splitting and channel prefixing variables. We showed that in cyclic shift symmetric wiretap channels, it suffices to solve an optimization problem only over one auxiliary random variable. We explicitly characterized the optimal rate splitting and channel prefixing random variables in terms of the solution of this single variable optimization problem. Next, we provided a sufficient condition for rate splitting to be unnecessary for achieving the boundary of the rate-equivocation region in cyclic shift symmetric wiretap channels; we showed that when $I(X;Y)-I(X;Z)$ is maximized at the uniform input distribution, i.e., when the wiretap channel is dominantly cyclic shift symmetric, then rate splitting is unnecessary. We also showed that when the wiretap channel is more capable and cyclic shift symmetric, $(C_B,C_s)$ rate-equivocation pair is achieveable. Finally, we applied our results to binary-input cyclic shift symmetric wiretap channels and characterized the boundaries of the rate-equivocation regions of the BSC-BEC and BEC-BSC wiretap channels. We provided full characterizations of the boundaries of the rate-equivocation regions for these wiretap channels by using a geometric framework. We found that rate-splitting is not necessary for the BSC-BEC wiretap channel. We also showed the existence of more capable, not less noisy, dominantly cyclic shift symmetric wiretap channels. 

\appendix

\appendixpage

\section{Proof of Theorem \ref{tm_mc}} \label{app2}
Assume that $p(y|x)$ is more capable than $p(z|x)$. For any $U
\rightarrow V \rightarrow X \rightarrow Y,Z$ and $\mu>0$, we
have
\begin{align}
&\mu  I(V;Y) + I(V;Y|U) - I(V;Z|U)\nonumber \\ \label{www} & = \mu [I(X;Y) - I(X;Y|V)] + I(X;Y|U) - I(X;Z|U) - [I(X;Y|V,U) - I(X;Z|V,U)]  \\ & = \mu I(X;Y) + I (X;Y|U) - I(X;Z|U) - [(\mu +1)I(X;Y|V) - I(X;Z|V)] \label{xxx} \\
& \leq \mu I(X;Y) + I(X;Y|U) - I(X;Z|U) \label{sss}
\end{align}
where (\ref{www}) and (\ref{xxx}) follow from the Markov chain
$U \rightarrow V \rightarrow X \rightarrow Y,Z$, and (\ref{sss})
follows from the more capable condition. Therefore, using a
non-trivial channel prefixing yields a loss in the objective
function $\mu I(V;Y) + I(V;Y|U)-I(V;Z|U)$ and $V=X$ is the
optimal selection. In other words, in order to characterize the entire rate-equivocation region it suffices to solve the following optimization problem: 
\begin{align} \label{oprob} \max_{U\rightarrow X \rightarrow Y,Z} \mu I(X;Y) + I(X;Y|U) - I(X;Z|U) \end{align} 
We claim that $|\mathcal{U}| \leq |\mathcal{X}|$ is sufficient for the solution of (\ref{oprob}). Given $U \rightarrow X \rightarrow Y,Z$, we fix the following $|\mathcal{X}|$ continuous functions of $P_{X|U}(x|u)$: $|\mathcal{X}|-1$ components of $P_{X|U}(x|u)$, $x=1,\ldots,|\mathcal{X}|-1$ and $I(X;Y|U=u) - I(X;Z|U=u)$. By Lemma 3 in \cite{ahlswede75} and the strengthened Caretheodory theorem of Fenchel-Eggleston in \cite{salehi78}, there exists a random variable $U^{'}\rightarrow X \rightarrow Y,Z$ such that \begin{align} \mu I(X;Y) + I(X;Y|U) - I(X;Z|U) = \mu I(X;Y) + I(X;Y|U^{'}) - I(X;Z|U^{'}) \end{align} Therefore, the optimization problem in (\ref{oprob}) can be solved with the cardinality bound $|\mathcal{U}|\leq |\mathcal{X}|$. Note the equivalence of the operations performed for proving the cardinality bounds in (\ref{oprob}) and in (\ref{ck3}).

To prove the converse statement, assume that $p(y|x)$ is not more capable than $p(z|x)$ and $f(P_x)$ is maximized at an interior point of $\Delta$. That is, $f(P_x^*)\geq f(P_x)$, $\forall P_x \in \Delta$, for some $P_x^*$ that has all non-zero entries. Moreover, as the more capable condition does not hold, $f(\hat{P}_x)<0$ for some input distribution $\hat{P}_x$. We use $\hat{P}_x$ and $P_x^*$ to construct a non-trivial channel prefixing $V$ such that $ V \rightarrow X \rightarrow Y,Z$ and
\begin{align}
f(P_x^*) <  I(V;Y)-I(V;Z) \label{yyy}
\end{align}
and hence show the existence of a non-trivial channel prefixing that provides a higher secrecy capacity and therefore a larger rate-equivocation region. Applying Lemma~\ref{lm1} to the
distributions $\hat{P}_x$ and $P_x^*$, there exists
a PMF $\bm q \in \Delta$, with $q_1>0$ such that
\begin{align}\label{thrt}
P_x^* = q_{1} \hat{P}_x + \sum_{k=1}^{|\mathcal{X}|-1}q_{k+1}\ev_{j_{k}}
\end{align}
for some index set $J \subset \{1,\ldots,|\mathcal{X}|\}$ with
$|J|=|\mathcal{X}|-1$, and $j_{k} \in J$. We construct $V$
with $|\mathcal{V}|=|\mathcal{X}|$ in the following manner:
\begin{align}
p_{V}(v_k)&=q_{k}, \qquad k=1,\ldots,|\mathcal{X}|
\end{align}
In addition, we select
$p_{X|V}(x|v_1)=\hat{P}_x$, $p_{X|V}(x|v_2)=\ev_{j_{1}}$, \ldots,
$p_{X|V}(x|v_{|\mathcal{X}|})=\ev_{j_{|\mathcal{X}|-1}}$. Evaluating
$P_{x}=\sum_{k=1}^{|\mathcal{X}|}p_{V}(v_k)p_{X|V}(x|v_k)$, we
observe that, by (\ref{thrt}), the constructed $P_{x}$ and
the maximizer $P_x^*$ are the same. 
However, $I(X;Y|V) - I(X;Z|V)<0$ because given $V=v_{1}$,
\begin{align}
I(X;Y|V=v_{1}) - I(X;Z|V=v_{1})=f(\hat{P}_x)<0
\end{align}
while given $V=v_k$ for $k \neq 1$, \begin{align}I(X;Y|V=v_{k}) - I(X;Z|V=v_{k})=0 \end{align} As $q_1>0$, we have
\begin{align}
I(X;Y|V) - I(X;Z|V)<0 \label{zzz}
\end{align}
We take $\mu=0$ in (\ref{xxx}) and the problem reduces to the calculation of the secrecy capacity. Therefore, $U= \phi$ is optimal in this case due to \cite{CK78}. Then, using (\ref{zzz}) in (\ref{xxx}) for $\mu=0$ gives  \begin{align} \label{g9o} I(X;Y) - I(X;Z) < I(V;Y) - I(V;Z) \end{align}
which yields (\ref{yyy}), the desired result, since $P_x$ is $P_x^*$ and the left hand side of (\ref{g9o}) is $f(P_x^*)$.

\section{Proof of Theorem \ref{cyc}} \label{app3}

For given $\mu \geq 0$, the optimal selections $U^*$ and $V^*$ are the solutions of the following optimization problem:
\begin{align} \label{xyd} \max_{U \rightarrow V \rightarrow X \rightarrow Y,Z} \mu I(V;Y) + I(V;Y|U)-I(V;Z|U) \end{align}
By using the steps in (\ref{www})-(\ref{xxx}), we obtain an equivalent statement for (\ref{xyd}) as: 
\begin{align} \label{xyd2} \max_{U \rightarrow V \rightarrow X \rightarrow Y,Z} \mu I(X;Y) + I(X;Y|U) - I(X;Z|U) - [(\mu+1)I(X;Y|V)-I(X;Z|V)] \end{align}
We have the following bound for the objective function in (\ref{xyd2}): \begin{align} \nonumber & \mu I(X;Y) + I(X;Y|U) - I(X;Z|U) - [(\mu+1)I(X;Y|V)-I(X;Z|V)] \\ &\leq \max_{P_x} \mu I(X;Y) + \max_{U \rightarrow V \rightarrow X \rightarrow Y,Z} I(X;Y|U) - I(X;Z|U) - [(\mu + 1)I(X;Y|V) - I(X;Z|V)] \\ & \leq \mu I_{\uv}(X;Y) + \max_{\hat{V} \rightarrow X \rightarrow Y,Z} I(X;Y) - I(X;Z) - [(\mu + 1)I(X;Y|\hat{V}) - I(X;Z|\hat{V})] \label{ineq} \end{align} where $\uv$ denotes the $|\mathcal{X}|$ dimensional discrete uniform random variable, and $I_{\uv}(X;Y)$ denotes the mutual information obtained by choosing the PMF of $X$ as $\uv$. In (\ref{ineq}), we used the fact that $\max_{P_x} I(X;Y) = I_{\uv}(X;Y)$ as Bob's channel is cyclic shift symmetric. Moreover, we used the fact that $U$ is not needed, i.e., $U=\phi$, for maximizing $I(X;Y|U) - I(X;Z|U) - [(\mu + 1)I(X;Y|V) - I(X;Z|V)]$. Because, for given $U \rightarrow V \rightarrow X \rightarrow Y,Z$, we can always pick $u_i \in \mathcal{U}$ that maximizes \begin{align} \nonumber \max_{u_i \in \mathcal{U}}\ &I(X;Y|U=u_i) - I(X;Z|U=u_i) \\ \label{sxf} & - \sum_{v \in \mathcal{V}}[(\mu+1)I(X;Y|V=v)-I(X;Z|V=v)]p(V=v|U=u_i) \end{align} and therefore, choose a deterministic $U$ with $U=u^*$, where $u^*$ is the argument of the maximization in (\ref{sxf}). Consequently, we have
\begin{align} \nonumber \max_{U \rightarrow V \rightarrow X \rightarrow Y,Z} &I(X;Y|U) - I(X;Z|U) - [(\mu + 1)I(X;Y|V) - I(X;Z|V)] \\ &= \max_{\hat{V} \rightarrow X \rightarrow Y,Z} I(X;Y) - I(X;Z) - [(\mu + 1)I(X;Y|\hat{V}) - I(X;Z|\hat{V})] \label{lls}\end{align} Note that the right hand side of (\ref{lls}) is the claimed auxiliary optimization problem in the statement of the theorem. We use $\hat{V}$ notation to emphasize that the auxiliary random variables on the right and left hand sides of (\ref{lls}) are different. 

Next, we will show that the bound in (\ref{ineq}) is satisfied with equality for any cyclic shift symmetric wiretap channel. Let $\hat{V}$ with $p(\hat{V}=v)$ and $p(X|\hat{V}=v)$, $v \in \mathcal{\hat{V}}$, be the solution of the auxiliary problem in (\ref{lls}). First, we note that it suffices to consider $\hat{V}$ such that $|\mathcal{\hat{V}}|\leq |\mathcal{X}|$. This follows by the arguments we have used in the previous cardinality bound proofs. In particular, given $\hat{V} \rightarrow X \rightarrow Y,Z$, we fix $|\mathcal{X}|-1$ components of $P_{X|\hat{V}}(x|\hat{v})$, $j=1,\ldots,|\mathcal{X}|-1$, together with $(\mu + 1)I(X;Y|\hat{V}=\hat{v})-I(X;Z|\hat{V}=\hat{v})$. By Lemma 3 in \cite{ahlswede75} and the strengthened Caretheodory theorem of Fenchel-Eggleston in \cite{salehi78}, the problem in (\ref{lls}) can be solved with the cardinality bound $|\mathcal{\hat{V}}|\leq |\mathcal{X}|$. Note the equivalence of the operations performed for proving the bounds in this problem and those in (\ref{oprob}) and (\ref{ck3}).     

Now, we construct the optimal $U^*, V^*$ by using the optimal $\hat{V}$ for the auxiliary problem in (\ref{lls}) as in the statement of the theorem. In particular, we select the cardinalities as $|\mathcal{U}^*|=|\mathcal{X}|$ and $|\mathcal{V}^*|=|\mathcal{X}|^2$ with the distributions $p(U^*=u)=\frac{1}{|\mathcal{X}|}$ for $u \in \mathcal{U}^*$ and 
\begin{align} \label{ilk} p(V^*=v|U^*=1)&=p(\hat{V}=v),\qquad v \in \{1,\ldots,|\mathcal{X}|\} \\ p(V^*=v|U^*=1)&=0,\qquad v \notin \{1,\ldots,|\mathcal{X}|\} \\ p(x|V^*=v)&=p(x|\hat{V}=v),\qquad v \in \{1,\ldots,|\mathcal{X}|\}  \\ p(V^*=|\mathcal{X}|+v|U^*=2)&=p(\hat{V}=v), \qquad v \in \{1,\ldots,|\mathcal{X}|\} \\ p(V^*=v|U^*=2)&=0,\qquad v \notin \{|\mathcal{X}|+1,\ldots,2|\mathcal{X}|\} \\ p(x|V^*=|\mathcal{X}|+v)&=p(x|\hat{V}=v)(1), \qquad v \in \{1,\ldots,|\mathcal{X}|\}  \\ \nonumber & \ \vdots \\ p(V^*=(|\mathcal{X}|-1)|\mathcal{X}|+v|U^*=|\mathcal{X}|-1)&=p(\hat{V}=v), \qquad v \in \{1,\ldots,|\mathcal{X}|\} \\ p(V^*=v|U^*=|\mathcal{X}|)&=0,\qquad v \notin \{(|\mathcal{X}|-1)|\mathcal{X}|+1,\ldots,|\mathcal{X}|^2\} \\ p(x|V^*=(|\mathcal{X}|-1)|\mathcal{X}|+v)&=p(x|\hat{V}=v)(|\mathcal{X}|-1) \label{son} \end{align} The structure of the construction in (\ref{ilk})-(\ref{son}) is an expression of the $U^*\rightarrow V^* \rightarrow X^*$ in Fig. \ref{uvx}. Each element of $U^*$ generates the optimizing selection $p(\hat{V})$ for the problem in (\ref{lls}) over disjoint $|\mathcal{X}|$ elements of $V^*$. Each disjoint $|\mathcal{X}|$ element of $V^*$ generates cyclic shifts of the optimizing selection $p(X|\hat{V})$ for the input $X$. In (\ref{ilk})-(\ref{son}), we denote the $k$th cyclic shift of the conditional PMF for the channel input $X$, $p(x|\hat{V}=v)$, as $p(x|\hat{V}=v)(k)$. Note that the cardinality of $\hat{V}$ is $|\mathcal{X}|$ while that of the optimum $V^*$ is $|\mathcal{X}|^2$ and $|\mathcal{X}|^2$ conditional input PMFs, $p(x|V^*=v)$, are obtained by cyclic shifts of $|\mathcal{X}|$ conditional input PMFs, $p(x|\hat{V}=v)$.  

We first observe that $p(x|U^*=i)$ are cyclic shifts of a fixed PMF over $X$ for different $i$. In particular, in the construction in (\ref{ilk})-(\ref{son}), we selected $p(x|V^*=v)$ as cyclic shifts of $p(x|\hat{V}=v)$ while we kept $p(V^*=v)$ the same as $p(\hat{V}=v)$. Hence, we have \begin{align} p(x|U^*=i) = p_f(x)(i-1) \end{align} where $p_f(x)=\sum_{v=1}^{|\mathcal{X}|} p(x|\hat{V}=v)p(\hat{V}=v)$. Note that $p_f(x)$ is the maximizing input PMF for the auxiliary problem. Therefore, $U^*$ and $V^*$ generate a uniform PMF for $X$: \begin{align}\label{eee} p(x) = \sum_{i=1}^{|\mathcal{X}|} p(U^*=i)p(x|U^*=i) = \sum_{i=1}^{|\mathcal{X}|}\frac{1}{|\mathcal{X}|}p_f(x)(i-1) = \frac{1}{|\mathcal{X}|} \end{align} 

Moreover, by construction of $U^*$ and $V^*$ and the cyclic shift symmetry of the channels, we observe that, for any given $i$, \begin{align} \nonumber \sum_{v=1}^{|\mathcal{X}|} [(\mu+1)I(X;Y|V^*=v+(i-1)|\mathcal{X}|) - &I(X;Z|V^*=v+(i-1)|\mathcal{X}|)]\\ &p(V^*=(i-1)|\mathcal{X}|+v|U=i) \nonumber \\  = \sum_{v=1}^{|\mathcal{X}|} [(\mu+1)I(X;Y|\hat{V}=v) - I(X;Z&|\hat{V}=v)]p(\hat{V}=v) \\ = (\mu+1)I(X;Y|\hat{V}) - I(X;Z|\hat{V}) \hspace{0.35in}& \label{usng}\end{align} Therefore, we have \begin{align} \nonumber &(\mu+1)I(X;Y|V^*) - I(X;Z|V^*) \\ &= \sum_{v = 1}^{|\mathcal{X}|^2} \Big((\mu+1)I(X;Y|V^*=v) - I(X;Z|V^*=v)\Big)p(V^*=v) \\ &= \sum_{i=1}^{|\mathcal{X}|}\sum_{v=1}^{|\mathcal{X}|^2} \Big((\mu +1)I(X;Y|V^*=v) - I(X;Z|V^*=v)\Big)p(v|U^*=i)p(U^*=i) \\ &= \frac{1}{|\mathcal{X}|} \sum_{i=1}^{|\mathcal{X}|}\sum_{v=1}^{|\mathcal{X}|}\Big( (\mu+1)I(X;Y|V^*=v+(i-1)|\mathcal{X}|) - I(X;Z|V^*=v+(i-1)|\mathcal{X}|)\Big)p(\hat{V}=v) \\ \label{eto} &= (\mu+1)I(X;Y|\hat{V}) - I(X;Z|\hat{V}) \end{align} where (\ref{eto}) is obtained by using (\ref{usng}) and the fact that $p(v|U^*=i)$ is non-zero only for $(i-1)|\mathcal{X}|+1 \leq v \leq i|\mathcal{X}|$. Note that \begin{align} I(X;Y|U^*=i) - I(X;Z|U^*=i) &= f(p_f(x)(i-1)) \\ &= f(p_f(x)), \qquad \forall i \end{align} Hence, given $U^* = i$, we have \begin{align} \nonumber & \hspace{-0.5in} I(X;Y|U^*=i) - I(X;Z|U^*=i) - \left[(\mu + 1)I(X;Y|V^*) - I(X;Z|V^*)\right] \\ &= f(p_f(x)) - \left[(\mu + 1)I(X;Y|\hat{V}) - I(X;Z|\hat{V})\right] \end{align} As $p_f(x)$ is the maximizing input PMF for the auxiliary problem, we have \begin{align} \nonumber \hspace{-0.4in}I(X&;Y|U^*) - I(X;Z|U^*) -[(\mu+1)I(X;Y|V^*) - I(X;Z|V^*)] \\ &= \max_{\hat{V} \rightarrow X \rightarrow Y,Z} I(X;Y) - I(X;Z) - [(\mu+1)I(X;Y|\hat{V}) - I(X;Z|\hat{V})] \label{achb} \end{align} Since $U^*$ and $V^*$ generate a uniform PMF for $X$ by (\ref{eee}), $I(X;Y)$ achieves its maximum, as well. Combining this with (\ref{achb}), we conclude that the constructed $U^*$ and $V^*$ achieve the upper bound in (\ref{ineq}) and hence are optimal.

\section{Proof of Lemma \ref{ism}} \label{app4}

Let the optimal selection of the auxiliary $\hat{V}$ in the following problem be $\hat{V}^*$: 
\begin{align} \label{sst} \max_{\hat{V} \rightarrow X \rightarrow Y,Z} I(X;Y)-I(X;Z)-[(\mu+1)I(X;Y|\hat{V})-I(X;Z|\hat{V})] \end{align} 
We will prove that at least one such $\hat{V}^*$ satisfies the following property: 
\begin{align} p(\hat{V}^*=v_k)=\frac{1}{|\mathcal{X}|}, \quad \mbox{and} \quad p(X=x|\hat{V}^*=v_{k})=p(X=x|\hat{V}^*=v_1)(k-1), \quad \forall k  \label{prr}\end{align} 
Due to Theorem \ref{cyc}, we already know that the cardinality of $\hat{V}$ is bounded by $|\mathcal{X}|$.

Let $R_e^* = \max_{P_x} f(P_x)$. First, we obtain an upper
bound for the objective function in (\ref{sst}):
\begin{align}\nonumber
&\hspace{-1.0in}I(X;Y) - I(X;Z) - [(\mu+1)I(X;Y|\hat{V}) - I(X;Z|\hat{V})] \\
&\hspace{-0.7in}\leq  R_e^* - [(\mu + 1)I(X;Y|\hat{V}) - I(X;Z|\hat{V})] \label{ff} \\
&\hspace{-0.7in}\leq  R_e^* - \min_{P_x}[(\mu+1)I(X;Y) - I(X;Z)] \label{tt}
\end{align}
where (\ref{ff}) follows from $I(X;Y)-I(X;Z) \leq R_e^*$
and (\ref{tt}) is obtained by replacing $[(\mu+1) I(X;Y|\hat{V}) - I(X;Z|\hat{V})]$ with its
minimum possible value.

Now, we will show that the upper bound in (\ref{tt}) is
achieved by an auxiliary $\hat{V}^*$ of cardinality $|\mathcal{\hat{V}}|\leq |\mathcal{X}|$ with the desired property in (\ref{prr}). By the hypothesis,
$\uv=\arg\max_{P_x} I(X;Y) = \arg\max_{P_x} f(P_x)$.
Moreover, let $P_x^*=\arg\min_{P_x} [(\mu +1)I(X;Y) - I(X;Z)]$. Note that $P_x^*$ is different from the uniform distribution. 
By cyclic shift symmetry, there exist
$|\mathcal{X}|-1$ other input distributions that minimize
$(\mu +1)I(X;Y) - I(X;Z)$, which are cyclic shifts of $P_x^*$,
denoted by $P_x^*(i)$ for $i=1,\ldots,|\mathcal{X}|-1$. Therefore, we define
the channel prefixing $\hat{V}^*$ with $|\mathcal{\hat{V}}^*|=|\mathcal{X}|$ as
$p(\hat{V}^*=v_i)=1/|\mathcal{X}|$ with transition probabilities
$p(x|\hat{V}^*=v_1)=P_x^*$, $p(x|\hat{V}^*=v_2)=P_x^*(1)$, \ldots, and
$p(x|\hat{V}^*=v_{|\mathcal{X}|})=P_x^*(|\mathcal{X}|-1)$. Then, the input distribution is
$P_x=\sum_{i=1}^{|\mathcal{X}|}p(\hat{V}^*=v_i)P_x^*(i)=\textbf{u}$. For this selection
of $\hat{V}^*$, we have
\begin{align}
 &\hspace{-1.2in}I(X;Y) - I(X;Z) - [(\mu+1)I(X;Y|\hat{V}^*) - I(X;Z|\hat{V}^*)] \nonumber \\
&\hspace{-0.9in}=f(P_x = \uv) - \min_{P_x} [(\mu+1)I(X;Y) - I(X;Z)]  \\
&\hspace{-0.9in}=R_e^* - \min_{P_x}[(\mu+1)I(X;Y)-I(X;Z)] \label{dd}
\end{align}
Note that (\ref{dd}) is equivalent to the upper bound in
(\ref{tt}). Moreover, the specified channel prefixing $\hat{V}^*$ satisfies the desired property in (\ref{prr}) by construction.

\section{Proof of Corollary \ref{th_mc}} \label{appd}

First, we select $V^*=X$ due to Theorem~\ref{tm_mc}. Next, we
note that there exists at least one input distribution, denoted
by $P_x^*$, that maximizes $f(P_x)$, since it is a bounded
continuous functional of $P_x$ and the probability simplex
$\Delta$ is compact.

There exist $|\mathcal{X}|-1$ other input
distributions (cyclic shifts of $P_x^*$) that achieve the
maximum $f(P_x)$. Let us define the auxiliary $U$, with
$\mathcal{U}=\{u_1,\ldots,u_{|\mathcal{X}|}\}$, with marginal distribution
$p_U(u_i)=\frac{1}{|\mathcal{X}|}$, and transition probabilities
$p_{X|U}(x|u_1)=P_x^*$, $p_{X|U}(x|u_2)=P_x^*(1)$, \ldots, and
$p_{X|U}(x|u_{|\mathcal{X}|})=P_x^*(|\mathcal{X}|-1)$, where $P_x^*(i)$ denotes the $i$th
cyclic shift of $P_x^*$.

Evaluating (\ref{ck2}) with the specified choice of $U^*$ and
with $V^*=X$, we have $I(V^*;Y)=C_B$, since $P_x=\sum_{u \in
\mathcal{U}} p_U(u)p_{X|U}(x|u) = \frac{1}{|\mathcal{X}|}\sum_{i=1}^{|\mathcal{X}|}
P_x^*(i)=\uv$, where $\uv$ is the uniform distribution, and since Bob's channel is cyclic shift symmetric. On
the other hand, evaluating (\ref{ck1}) for this specific
choice, we get $I(X;Y|U) - I(X;Z|U) = C_s$, since for any $u
\in \mathcal{U}$, $I(X;Y|U=u) - I(X;Z|U=u) = \max_{P_x} f(P_x)
= C_s$. This proves that $(C_B,C_s)$ pair is achievable.

Note that if $P_x^* = \uv$, i.e., if the channel satisfies dominant cyclic shift symmetry, then $U^*=\phi$ is optimal since any cyclic shift of $\uv$ is $\uv$ itself and thus $U^*$ is independent of $X$.

\bibliographystyle{unsrt}

\end{document}